\documentclass[fleqn,usenatbib]{mnras}

\usepackage{newtxtext,newtxmath}

\usepackage[T1]{fontenc}
\usepackage{ae,aecompl}

\usepackage{amsmath}	
\usepackage{amssymb}	
\usepackage{adjustbox}
\usepackage{pdflscape}
\usepackage{graphicx}

\usepackage{hyperref}	
\usepackage{cleveref}

\hypersetup{colorlinks=true,linkcolor=blue,citecolor=blue,filecolor=blue,urlcolor=blue}

\renewcommand{\theenumi}{\arabic{enumi}.}
\renewcommand{\theenumii}{\theenumi \arabic{enumii}}

\title[The CGM and IGM at z$\sim$5: metal budget and physical connection]{The CGM and IGM at z$\sim$5: metal budget and physical connection}

\author[A. Codoreanu et al]{
Alex Codoreanu,$^{1,2,3}$\thanks{E-mail: acodoreanu@swin.edu.au}
Emma V. Ryan-Weber,$^{1,2}$
Luz {\'A}ngela Garc{\'i}a,$^{1,2,4}$
\newauthor
Neil H.M. Crighton,$^{1}$
George Becker,$^{5}$ 
Max Pettini,$^{6}$
Piero Madau,$^{7}$
and Bram Venemans$^{8}$
\\
$^{1}$Centre for Astrophysics and Supercomputing, Swinburne University of Technology, Hawthorn, Victoria 3122, Australia\\
$^{2}$ARC Centre of Excellence for All-sky Astrophysics (CAASTRO)\\
$^{3}$Centre for Transformative Innovation \\
$^{4}$Universidad ECCI, Carrera 19 No. 49 - 20. Bogot{\'a}, Colombia\\
$^{5}$Department of Physics Astronomy, University of California, Riverside, 900 University Avenue, Riverside, CA 92521, USA\\
$^{6}$Institute of Astronomy, Madingley Road, Cambridge, CB3 0HA, UK\\
$^{7}$Department of Astronomy $\&$ Astrophysics, University of California, 1156 High Street, Santa Cruz, CA 95064, US \\
$^{8}$Max-Planck Institute for Astronomy, K$\ddot{o}$nigstuhl 17, D-69117 Heidelberg, Germany \\
}

\date{Accepted XXX. Received YYY; in original form ZZZ}

\pubyear{2017}

\begin{document}
\label{firstpage}
\pagerange{\pageref{firstpage}--\pageref{lastpage}}
\maketitle

\begin{abstract} 
We present further results of a survey for absorption line systems in the spectra of four high redshift quasars (5.79 $\le$ z$_{\textrm{em}}$ $\le$ 6.13) obtained with the ESO Very Large Telescope X-Shooter. We identify 36 \ion{C}{IV} and 7 \ion{Si}{IV} {systems} with a $\ge$ 5$\sigma$ significance. The highest redshift \ion{C}{IV} and \ion{Si}{IV} absorbers identified in this work are at z = 5.80738 $\pm$ 0.00017 and z = 5.77495 $\pm$ 0.00038, respectively. We compute the comoving mass density of \ion{Si}{IV} ($\Omega_{\ion{Si}{IV}}$) and find that it evolves from $\Omega_{\ion{Si}{IV}}$ = 4.3$^{+2.1}_{-2.1}$ $\times$10$^{-9}$ at <z> = 5.05 to $\Omega_{\ion{Si}{IV}}$ = 1.4$^{+0.6}_{-0.4}$ $\times$10$^{-9}$ at <z> = 5.66. We also measure $\Omega_{\ion{C}{IV}}$ = 1.6$^{+0.4}_{-0.1}$ $\times$10$^{-8}$ at <z> = 4.77 and $\Omega_{\ion{C}{IV}}$ = 3.4$^{+1.6}_{-1.1}$ $\times$10$^{-9}$ at <z> = 5.66. We classify our \ion{C}{IV} absorber population by the presence of associated \textit{low} and/or \textit{high ionisation} systems and compute their velocity width ($\Delta$v$_{90}$). We find that all \ion{C}{IV} systems with $\Delta$v$_{90}$ > 200 kms$^{-1}$ have associated \textit{low ionisation} systems. We investigate two such systems {separated by 550 physical kpc along a line of sight,} and find it likely that they are {both} tracing a multi-phase medium where hot and cold gas is mixing at the interface between the CGM and IGM. We further discuss the \ion{Mg}{II} systems presented in {a previous work} and we identify 5 \ion{Si}{II}, 10 \ion{Al}{II}, 12 \ion{Fe}{II}, 1 \ion{C}{II}, 7 \ion{Mg}{I} and 1 \ion{Ca}{II} associated transitions. We compute the respective comoving mass densities in the redshift range 2 to 6, as allowed by the wavelength coverage.

\end{abstract}





\section{Introduction}

Absorption systems in the spectra of high redshift quasi-stelar objects (QSOs) present an opportunity to identify and study intervening metal enriched clouds during the first billion years of evolution of the Universe. {Unambigous detection of t}hese clouds {is provided} by the presence of doublets with high oscillator strength such as \ion{Mg}{II} \citep{KACPRZAK2011, KACPRZAK2011B, KACPRZAK2012a, MATEJECK2012, KACPRZAK2013, BOSMAN2017, CODOREANU2017}, \ion{Si}{IV} \citep{BOKSENBERG2015, SONGAILA2005} as well as \ion{C}{IV} \citep{SIMCOE2006, RYANWEBER2006, BECKER2009, RYANWEBER2009, SIMCOE2011, DODORICO2013, GONZALO}. The rest frame ionisation wavelength and oscillator strength of each transition identified in this work are presented in Table \ref{tab:iontable}.

These absorption systems are generally categorised into \textit{low} and \textit{high ionisation} systems. \textit{Low ionisation} absorption systems (\ion{Mg}{II}, \ion{Si}{II}, \ion{Al}{II}, \ion{O}{I} and others) trace low temperature/high density regions connected to the circumgalactic medium (CGM) (eg. \citealt{STEIDEL2010, NIELSEN2013a, NIELSEN2013b, CHURCHILL2013a, NIELSEN2015, NIELSEN2016}) while \textit{high ionisation} (\ion{C}{IV}, \ion{Si}{IV} and others) trace high temperature/low density regions generally associated with the intergalactic medium (IGM) (eg. \citealt{SCHAYE2003, SCHAYE2007}; \citealt{SCHAYE2004}, \citealt{DODORICO2016}; \citealt{FINLATOR2016}; \citealt{KEATING2016}; \citealt{GARCIA2017}; \citealt{OPPENHEIMER2017}).

However, such associations and boundaries are not always applicable as \ion{C}{IV} systems with column densities log(N/cm$^{2}$)$>$13.5 have also been associated with the halos of galaxies with stellar mass (M$_{\ast}$) >10$^{9.5}$M$_{\odot}$ at 0.0015 $<${\it z}$<$ 0.015 \citep{BURCHETT2016}. {Furthermore, \cite{STEIDEL2010} and \cite{TURNER2014} identify and measure the optical depth of \ion{C}{IV} systems and find that they are preferentially found at a proper transverse distance of less than 200kpc of {\it z}$\simeq$2-3 galaxies.}. {\cite{ADELBERGER2005} also connects \ion{C}{IV} systems with column densities log(N/cm$^{2}$)$>$14.0 to young star forming field galaxies up to {\it z}$\approx$3.3 and finds that their gas halo can extend up to 80 kpc. Thus, {\it high ionisation} systems can also be found in virialised halos where they are physically mixed with {\it low ionisation} systems.}

Replicating the observed abundance and evolution of {absorption} systems {beyond} redshift 5 is a challenging task as they trace different physical environments and are sensitive to a large number of connected and correlated physical processes. For example, the galaxy contribution to the shape and amplitude of the global UV background (UVB, \citealt{OPPENHEIMER2009}) {beyond redshift 5} depends on both the volume density of galaxies \citep{ATEK2015a, MASON2016, LIVERMORE2017, BOUWENS2015, BOUWENS2017b} {and the properties of both Pop II stars and the first metal-free stars (PopIII) (i.e. \citealt{HEGER2002, YOSHIDA2006, HOSOKAWA2012, PALLOTINI2014}). The relative abundance patterns described in \cite{BECKER2012} are consistent with a scenario in which metal production in Pop II stars dominates the metal budget by z$\sim$6 but, currently, there is no clear observational tracer to signal the transition from PopIII to PopII stars which is expected to occur before z$\sim$10 \citep{MAIO2010}.}

Furthermore, absorption systems are also sensitive to the {yield and return fraction} of metals \citep{MADAU2014} which itself depends on both the initial mass function \citep{SALPETER1955, KROUPA2001, CHABRIER2003} and the outflow models which transport those metals from the interstellar medium (ISM) to the IGM through the CGM \citep{FERRARA2000, MADAU2001,OPPENHEIMER2006}. Absorption systems are then an important observational discriminant as they trace diverse temperature and density regions and, provide a census of metals \citep{WENLAN2017} which is not limited by the brightness of the galaxies associated with the enrichment.

Recent works by \citet{BOSMAN2017} (B17) and \citet{CODOREANU2017} (C17) have identified a population of \textit{weak} \ion{Mg}{II} systems (W$_{2796}$\footnote{The equivalent width of the \ion{Mg}{II} $\lambda$2796 transition } $\le$ 0.3 \AA) from redshift 5 to 7. Previous works by \citet{BECKER2006, BECKER2011} have also identified \textit{low ionisation} absorbers with multiple associated transitions (eg. \ion{O}{I}, \ion{Si}{II}, \ion{C}{II}) beyond redshift 5. Their high incidence rates suggest that they are most likely tracing the small and numerous galaxies needed to reionise the Universe during the Epoch of Reionisation (EoR) with M$_{\textrm{UV}}$ $\le$ -13 \citep{ROBERTSON2013}. While these absorbers have not yet directly been connected to specific stellar populations, their presence indicates that metals, as traced by \textit{low ionisation} systems, have already been well established and have a significant cross-section by redshift $\sim$6.

While these \textit{weak} \ion{Mg}{II} systems have a significant cross-section, they do not {account for} a large fraction of the metal budget. C17 has shown that \ion{Mg}{II} systems with W$_{2796}$ $\le$ 1 \AA $ $ hold a small fraction ($\sim$1/50) of the comoving mass density of \ion{Mg}{II} ($\Omega_{\ion{Mg}{II}}$) in the redshift range 4.03$<$z$\le$ 5.45. Interestingly, C17 has also shown that $\Omega_{\ion{Mg}{II}}$, as measured by all systems discovered in their work\footnote{0.117 $\le$ W$_{2796}$ $\le$ 3.655 \AA $ $ }, increases from $\Omega_{\ion{Mg}{II}}$ = 2.1$^{+6.3}_{-0.6} \times$10$^{-8}$ at <z> = 2.48 to $\Omega_{\ion{Mg}{II}}$ = 3.9$^{+7.1}_{-2.4} \times$10$^{-7}$ at <z> = 4.77.

This order of magnitude increase is in contrast to the evolution of the comoving mass density of \ion{C}{IV} ($\Omega_{\ion{C}{IV}}$) which has a flat evolution across a similar redshift range \citep{SONGAILA2001, PETTINI2003}. For example, \citet{BOKSENBERG2015} (B15) measure a mean $<$ $\Omega_{\ion{C}{IV}}$ $>$ = 1.23 $\pm$ 0.66$\times$10$^{-8}$ at a mean redshift $ < $z$>$ = 3.20 across the redshift range 1.9 $<$ z $<$ 4.5 and \cite{DODORICO2013} measure $\Omega_{\ion{C}{IV}}$ = 1.4 $\pm$ 0.3$\times$10$^{-8}$ at z = 4.818. However, from redshift 5 to 6, $\Omega_{\ion{C}{IV}}$ declines by a factor of 2 to 4 \citep{SIMCOE2006, SIMCOE2011, RYANWEBER2006, BECKER2009, RYANWEBER2009, DODORICO2013}.

Th{e evolution in $\Omega_{\ion{C}{IV}}$} can be driven either by a change in the ionisation state, the enrichment of the IGM \citep{BECKER2015} or both. It is reproduced in simulations by \citet{OPPENHEIMER2006}, \citet{OPPENHEIMER2009}, \citet{CEN2011} and \citet{GARCIA2017}.
However, the opposite evolution of $\Omega_{\ion{Mg}{II}}$ when compared to the evolution of $\Omega_{\ion{C}{IV}}$ does not support an increase in the metal budget as the primary driver in the evolution of $\Omega_{\ion{C}{IV}}$ (C17). This suggests that the evolution of $\Omega_{\ion{C}{IV}}$ is then mostly driven by {a change in the ionisation state of the IGM resulting from changes in the UVB beyond redshift 5}.

\citet{FINLATOR2016} investigate the impact of three different UVB prescriptions on the resulting absorber population and compare to the observational results of \cite{BECKER2011} and \cite{DODORICO2013} (D13). The three types of UVBs tested are (1) the UVB presented by \cite{HM12} (HM12), (2) a modified and directly simulated version of the HM12 UVB which accounts for inhomogeneous galaxy emissivity combined with a QSO emissivity, and (3) a QSO emissivity only. They find that while all three UVB scenarios can reproduce the column density distribution functions (CDDFs) of \ion{C}{II}, \ion{Si}{IV}, and \ion{C}{IV} only (2) reproduces the observed ionic ratios of \ion{Si}{IV}/\ion{C}{IV} and \ion{C}{II}/\ion{C}{IV}.

Interestingly, the QSO-only UVB model greatly overproduces photons with energies greater than 4 Ryd which is reflected in an over production of \ion{C}{IV} (4.7 Ryd). However, at lower energies, probed by \ion{Si}{IV}, \ion{Si}{II} and \ion{C}{II}, all three UVB models have similar intensities. Thus simply scaling up the intensity of the UVB would result in different {ratios of \ion{Si}{IV}/\ion{C}{IV} and \ion{C}{II}/\ion{C}{IV}} than a UVB which is adjusted by considering varying contributions from sources of harder photons such as QSOs or PopIII stars \citep{FINLATOR2016}. Recently, \cite{DOUGHTY2018} have also shown that aligned absorber pairs (eg. multiple ions associated with the same enriched and ionised gas halo) can be used to improve the constraints on the UVB. They found that the observed statistics of \ion{C}{IV}/\ion{Si}{IV} are best reproduced by a hard, spatially uniform UVB but, a single aligned \ion{Si}{II}/\ion{Si}{IV} pair is reproduced best by the HM12 UVB prescription.

Thus, understanding the evolution of both \textit{low} and \textit{high ionisation} systems is then necessary as \textit{low ionisation} systems (i.e. \ion{C}{II}, \ion{Si}{II}) are sensitive to both changes in ionisation and gas density/metallicity \citep{FINLATOR2015} while \ion{C}{IV} is mostly sensitive to ionisation prescriptions \citep{FINLATOR2016}. Future simulations can explore if there is a preferred balance between changes to a global UVB from local sources in conjunction with different prescriptions for self-shielded halos which produce metals but do not contribute photons to the global UVB. In order to test such scenarios several questions arise:

- ``How does $\Omega_{\ion{Si}{IV}}$ evolve beyond redshift 5?''

- ``What is the physical connection between \textit{low} and \textit{high ionisation} systems beyond redshift 5?''

-``Are \textit{low and high ionisation} systems tracing distinct physical constructs or are they tracing multiphase gas at virial distances or at the interface between the CGM and IGM?''

- ``What is the evolution of other species of \textit{low ionisation} systems (i.e. \ion{Si}{II}, \ion{C}{II}, \ion{Fe}{II} and others)?''

We explore these questions by searching for intervening absorption systems in four medium resolution and signal-to-noise spectra of redshift $\sim$6 QSOs. These spectra were investigated for the presence of \ion{Mg}{II} systems and the results were presented in C17. In this work, we provide the first $\Omega_{\ion{Si}{IV}}$ values and corresponding CDDFs {beyond redshift 5. We} compare with \ion{C}{IV} in the range 4.92<z<6.12. We also identify \ion{Si}{II}, \ion{Al}{II}, \ion{Fe}{II}, \ion{C}{II}, \ion{Mg}{I} and \ion{Ca}{II} associated transitions in addition to the \ion{Mg}{II} discussed in C17. We discuss the details of the identification of all absorption systems in Section \ref{sec:candidateselection}. We discuss our treatment of false positive contamination and completeness considerations in Section \ref{sec:surveycompandfp}. We provide the incidence rates, comoving mass densities and CDDFs (and best fit parameters) of the \ion{Si}{IV} and \ion{C}{IV} systems identified in this work in Section \ref{sec:absstats}. {We present the comoving mass densities of \ion{Mg}{I}, \ion{Ca}{II}, \ion{Si}{II}, \ion{Al}{II}, \ion{Fe}{II} and \ion{C}{II} and} discuss our results in Section \ref{sec:discussion}. We provide a summary and conclusions in Section \ref{sec:conclusion}. Throughout this paper we use a $\Lambda$CDM cosmology with $\Omega_{M}$ = 0.308 and $H_0$ = 67.8kms$^{-1}$Mpc$^{-1}$ \citep{PLANCK2015}.


\section{Candidate selection}
\label{sec:candidateselection}

The observations, exposure times, data reduction and instrument resolution are described in C17. The reduced spectra are binned with a resolution of 10 kms$^{-1}$ pixel$^{-1}$. We follow the same steps {as C17} in identifying absorbers and quantifying the completeness of our survey. In short, we first create a candidate list using an automatic search algorithm whose output is then visually inspected by the lead author (AC) and a final list of candidates is created. These candidate absorbers are then fit with Voigt profiles using \textsc{VPFIT 10.0} \citep{VPFIT}.

\subsection{Automatic Search}
\label{sec:autosearch} 

In order to identify \ion{C}{IV} and \ion{Si}{IV} doublets, we create a candidate list by finding all pixels of the spectra which meet the following conditions: \\

\noindent$\bullet$ a minimum 3 consecutive 5$\sigma_i$ pixel detections;

\hspace{16mm} $\sigma_i$ = $\frac{(1-F_{\lambda}i)}{E_{\lambda}}$

\noindent$\bullet$ $W_{d1}$/$\sigma$ $W_{d1}$ $\ge$ 5 or $W_{d2}$/$\sigma$ $W_{d2}$ $\ge$ 5

\noindent$\bullet$ 0.5$\le$($W_{d1}$ $\pm$ $\sigma$ $W_{d1}$)/($W_{d2}$ $\pm$ $\sigma$ $W_{d2}$)$\le$5, searching for \ion{C}{IV}

\noindent$\bullet$ 0.8$\le$($W_{d1}$ $\pm$ $\sigma$ $W_{d1}$)/($W_{d2}$ $\pm$ $\sigma$ $W_{d2}$)$\le$ 4, searching for \ion{Si}{IV} \\

\noindent where $F_{\lambda}i$ and $E_{\lambda}i$ are the flux, error values associated with a pixel $i$. $W$ and $\sigma W$ represent the equivalent width and error of the consecutive pixels associated with the \ion{C}{IV} $\lambda\lambda$1548 1550 and \ion{Si}{IV} $\lambda\lambda$1393 1402 doublet candidates where $d1$ and $d2$ denote each one of the respective candidates. The respective equivalent width ratios\footnote{($W_{d1}$ $\pm$ $\sigma W_{d1}$)/{($W_{d2}$ $\pm$ $\sigma W_{d2}$)}} are chosen by the lead author in order to minimise the contamination by false positives. {These ratios represent only the pixel flux and error values and do not account for possible blends with other transitions}. The selection algorithm outputs 94 candidate \ion{C}{IV} systems (with 118 components) and 18 candidate \ion{Si}{IV} systems (with 40 components).

\subsection{Visual check}
\label{sec:visualcheck} 

Each candidate is visually inspected by the lead author (AC) and selected as an absorber based on the similarity of the velocity profile of the two transitions in each doublet. In C17, most of the rejected \ion{Mg}{II} candidates occurred in the NIR where telluric absorption and sky-line emissions heavily polluted the spectra. In the present work, the majority of the absorption path {-redward of the Ly$\alpha$ emission peak of the QSO to within 3000 kms$^{-1}$ of the QSO redshift-} for \ion{C}{IV} and \ion{Si}{IV} is in the VIS arm of X-Shooter. 

\begin{figure}
 \includegraphics[width=8.4cm]{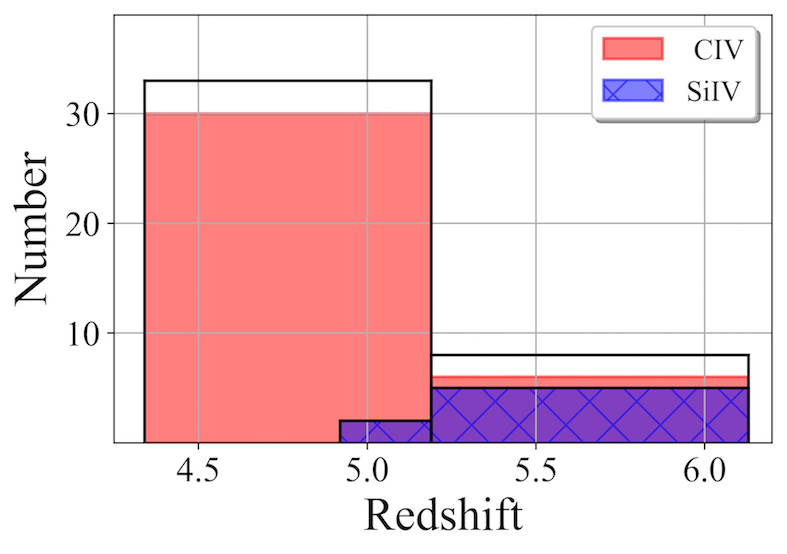}
 \caption{Histograms of \ion{C}{IV} and \ion{Si}{IV} discovered systems. The raw counts are denoted by black outlines while the number of systems which pass the 5$\sigma$ recovery selection criteria are shaded in red and cross-hatched blue respectively. All values are presented in Table \ref{tab:incidenceomegatable}.}
 \label{fig:c4histo}
\end{figure}

The rejected candidates have mis-matched velocity profiles or are weak features dominated by RMS fluctuations. From the 94 \ion{C}{IV} candidates, 41 are {accepted} and from the 18 \ion{Si}{IV} candidates, 7 are {accepted} by the lead author as possible 'true' absorbers (see Figure \ref{fig:c4histo} and $\bar{N}$ column in Table \ref{tab:incidenceomegatable}). Following this selection, we search for associated ions (\ion{C}{II}, \ion{Si}{II}, \ion{Mg}{I}, \ion{Al}{II}, \ion{Al}{III}, \ion{N}{V}, \ion{O}{I}, \ion{O}{VI}, \ion{Fe}{II} and \ion{Ca}{II}) by using the \ion{C}{IV}, \ion{Si}{IV} and \ion{Mg}{II} (described in detail in C17) doublets as redshift anchors for their possible location. For this task we use the \textsc{plotspec} package\footnote{developed by \href{https://github.com/nhmc/plotspec/}{Dr. Neil Crighton} \\ \href{https://github.com/nhmc/plotspec/}{https://github.com/nhmc/plotspec/}}.

We find 12 systems with multiple associated ions and each one is discussed in Sec. \ref{sec:sols}. The \ion{C}{IV} associated with system 9 in sightline $ULAS$ $J1319+0959$ does not meet our 5$\sigma$ discovery criteria (see section \ref{sec:autosearch}) but is included in our analysis for reasons discussed in sub-section \ref{sec:u1319}. It is the only absorber 'manually' introduced. The rejected systems are single component systems and have no other associated transitions (ie. \ion{Fe}{II}, \ion{C}{II}). Thus, we are not forced to consider multiple ions when trying to ascertain the veracity of an absorption system.

\begin{table}
 \centering
 \caption{Rest frame ionisation wavelength ($\lambda$) and oscillator strength ($f$) of the ions identified in this work as reported in \citet{MORTON2003}.}
 \label{tab:iontable}
 \begin{tabular}{|| c | c | c ||}
 \hline
 ion & $\lambda$ ( \AA) & $f$ \\

 \hline
 \ion{C}{IV} \hfill & \hfill 1548.2049 & \hfill 0.189900 \\ 
 & \hfill 1550.77845 & \hfill 0.094750 \\ 

 & & \\ 

 \ion{Si}{IV} \hfill& \hfill 1393.76018 & \hfill 0.513 \\ 
 & \hfill 1402.77291 & \hfill 0.254 \\ 
 
 & & \\ 

 \ion{Mg}{II} \hfill& \hfill 2796.3542699 & \hfill 0.6155 \\ 
 & \hfill 2803.5314853 & \hfill 0.3058 \\ 

 & & \\ 

 \ion{Mg}{I} \hfill & \hfill 2852.96328 & \hfill 1.830000 \\ 

 & & \\ 

 \ion{Al}{II}\hfill & \hfill 1670.7886 &\hfill 1.740 \\ 

 & & \\ 

 \ion{Si}{II}\hfill & \hfill 1526.70698 &\hfill 0.13300 \\ 
                    & \hfill1260.4221	& \hfill 1.180 \\
 & & \\ 

 \ion{C}{II} \hfill & \hfill 1334.5323 & \hfill 0.127800 \\ 

 & & \\ 

 \ion{Fe}{II}\hfill &\hfill 2600.1724835 & \hfill 0.2394 \\ 
 &\hfill 2586.6495659 & \hfill 0.069125 \\ 
 &\hfill 2382.7641781 &\hfill 0.320 \\ 
 &\hfill 2344.2129601 &\hfill 0.1142 \\ 
 &\hfill 1608.45085 & \hfill 0.0577 \\

 \end{tabular}
\end{table}

\subsection{Voigt Profiles \& Equivalent Widths}
\label{sec:vpew} 

We use \textsc{VPFIT 10.0} to fit Voigt profiles to the absorption lines {and we do not tie associated transitions together as they could trace multi-phase gas. However, \cite{BOKSENBERG2015} allow the fixed gas temperature to vary in the range 10$^{\text{4}}$<$T$<10$^{\text{5}}$K and has shown that the \textsc{VPFIT} retrieved column density parameters of Voigt profiles do not depend strongly on the temperature. For reference, the thermally broadened profile of a 10$^{\text{5}}$K gas cloud results in an upper bound for the thermal broadening parameter of $b_{therm}\simeq$10 km s$^{\text{-1}}$. As recommended in \cite{VPFIT}, we impose a minimum $b$ $<<$ $b_{expected}$. We choose a minimum value of $b$ = 1km/s.}

As the dominant source of uncertainty arises from the continuum fitting process, we adjust the continuum level by $\pm$ 5$\%$ and repeat the entire fitting procedure. This leads to the error bars associated with each set of Voigt profile parameters. The redshift ($z$) and Doppler parameter ($b$) of each absorber is fit individually and no absorbers reported in this work are highly saturated.

 A system is defined as all components within 0$\le\Delta${\it v}\footnote{$\Delta${\it v}=$c$ $\times \Delta \lambda$, where $c$ is the speed of light}$\le$500 kms$^{-1}$. We select this $\Delta v$ width range to account for the possible stellar velocity dispersion of galaxies with an intrinsic $B$ band magnitude M$_B \ge$-25 \citep{FABERJACKSON1976} as we have no {\it a priori} information on the associated galaxies. The velocity width of a system ($\Delta v_{\text{90}}$) is then computed from the corresponding wavelength boundary ($\Delta \lambda$; \citealt{PROCHASKA2008}) enclosing 90$\%$ of the optical depth of all components. We compute the equivalent width ($W_0$) of each component over each associated $\Delta \lambda$ value. The $W_0$ of blended systems is computed from the Voigt profile fit. The total column column density of an absorption system is computed by summing the column densities of the components.

{We present all identified systems, their components, equivalent width, Voigt profile parameters, associated errors and recovery levels in the system tables available in the online appendix}. All visually selected systems (see sec. \ref{sec:visualcheck}) from the automated output created by the detection algorithm (see sec. \ref{sec:autosearch}) are discussed in the following section.

\subsection{Individual sight lines}
\label{sec:sols} 

\begin{figure*}
	\includegraphics[width=14cm]{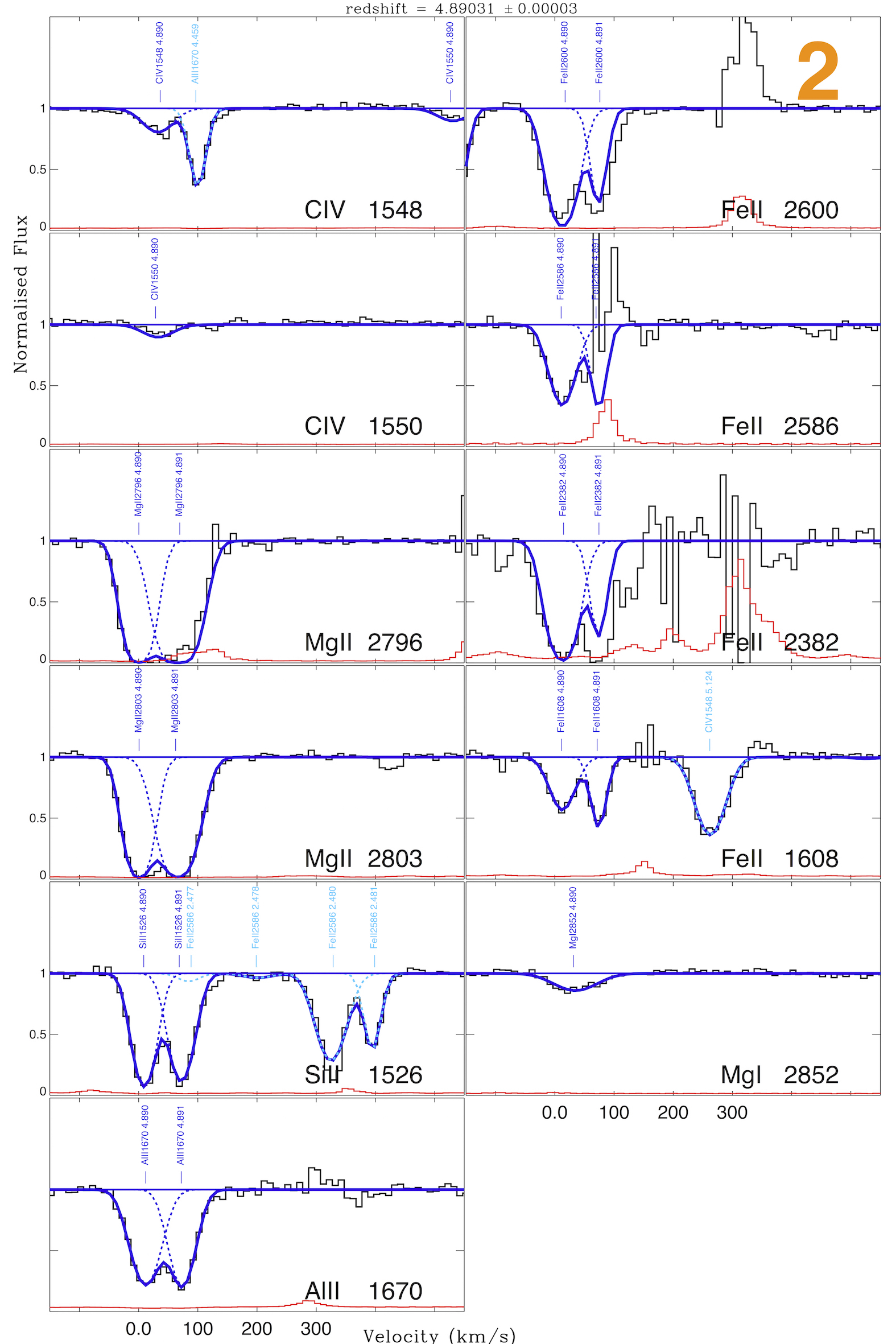}

	\caption{System 2 identified in the ULAS J0148+0600 sightline. Each transition is identified in the bottom right of each panel. In each panel, the vertical axis is the continuum normalised flux. The horizontal axis is the velocity separation ( kms$^{-1}$) from the lowest redshift component of a system. The normalised spectrum is plotted in black and the associated error is in red. The solid blue line represents the full fit to the spectra and includes other ions besides the transition identified in the bottom right of each panel. Individual components are plotted with dashed lines and are identified by a vertical label. The identified transition components are in solid blue and other transitions are in light blue.}	 
	\label{fig:U0148b}
\end{figure*}

{The four QSO redshifts, apparent magnitudes and initial discoveries are described in C17. We observed each object for $\sim$10 hours.} Below, we present all \ion{C}{IV} and \ion{Si}{IV} absorbers along with associated transitions. Given that all \ion{Mg}{II} absorbers have been presented in C17, we only include them in the system plots {available in the online appendix (see Fig. \ref{fig:U0148b} for an example)}. {No additional \ion{Mg}{II} absorbers are found after identifying the \ion{C}{IV} and \ion{Si}{IV} doublets}. Blended components are marked with a $\Downarrow$ while those polluted by a sky-line or poor subtraction residual are marked with a $\Uparrow$. Systems which do not meet the 5$\sigma$ recovery selection criteria in their respective (log(N), $dz_j$) bins are identified with a $^{\ast}$. We discuss this in detail in the follow-up Section \ref{sec:surveycompandfp}.

Additionally, we present, {in the online appendix}, the associated transitions (\ion{Mg}{I}, \ion{Fe}{II}, \ion{Al}{II} and \ion{Ca}{II}) of \ion{Mg}{II} discoveries from C17 with redshifts outside this paper's search region defined by the \ion{C}{IV} and \ion{Si}{IV} wavelengths. {For ease of understanding to those reading the following subsections in detail, we recommend to have the online appendix at hand. We only include Fig. \ref{fig:U0148b} and Table \ref{tab:u0148table} in the main paper as an example.}

\begin{table*}
 \centering
 \caption{Median redshift values (<z>), redshift bins ($\Delta z$), number of discovered systems ($\bar{N}$), number of recovery selected systems ($\ddot{N}$), adjustment scalars ($A$), redshift path ($dz$), comoving absorption path ($dX$), incidence rates ($dN/dz$), comoving incidence rates ($dN/dX$) and comoving mass densities ($\Omega$) for \ion{C}{IV} and \ion{Si}{IV}.}
 \label{tab:incidenceomegatable}
 \begin{tabular}{|| c | c | c | c | c | c | c | c | c | c | c ||}
 \hline
 ion & $ < $z$ > $ & $\Delta z$ & $\bar{N}$ & $\ddot{N}$ & $A$ & $dz$ & $dX$ & $dN/dz$ & $dN/dX$ & $\Omega$ \\

 \hline
 \ion{C}{IV}& 4.77 & 4.33-5.19 & 33 & 30 & 1.47 & 2.86 & 12.37 & 15.4 $\pm$ 2.8 & 3.6 $\pm$ 0.6 & 1.6$^{+0.4}_{-0.1}$ $\times$10$^{-8}$\\
 & 5.66 & 5.19-6.13 & 8 & 6 & 2.29 & 3.13 & 14.42 & 4.4 $\pm$ 1.2 & 0.9 $\pm$ 0.3 & 3.4$^{+1.6}_{-1.1}$ $\times$10$^{-9}$\\ \ion{Si}{IV}& 5.05 & 4.92-5.19 & 2 & 2 & 2.12 & 0.43 & 1.98 & 9.8 $\pm$ 4.7 & 2.2 $\pm$ 1.1 & 4.3$^{+2.1}_{-2.1}$ $\times$10$^{-9}$\\
 & 5.66 & 5.19-6.13 & 5 & 5 & 1.49 & 3.13 & 14.42 & 2.4 $\pm$ 0.9 & 0.5 $\pm$ 0.2 & 1.4$^{+0.6}_{-0.4}$ $\times$10$^{-9}$\\ 

 \end{tabular}
\end{table*}

\subsubsection{ULAS J0148+0600}
\label{sec:u0148} 
We present 9 new systems and the highest redshift absorber in this sightline which meets our 5$\sigma$ recovery selection criteria is system 9 with z = 5.82630 $\pm$ 0.00013. {All systems and associated components are presented in detail in Section A of the online Appendix.}

Systems 1, 5, 7 are single component \ion{C}{IV} systems. System 9 is a single \ion{Si}{IV} system with both the $\lambda\lambda$1393 1402 features blended with the \ion{C}{IV} $\lambda$1550 feature of system 7 and a \ion{Fe}{II}$\lambda$2382 absorber at z $\simeq$ 3.01858, respectively. {Furthermore, the \ion{Si}{IV} $\lambda$1393 could also be a \ion{Mg}{II} $\lambda$2796 system at z$\simeq$2.4024. This possible \ion{Mg}{II} system is not chosen from the initial candidate list by the lead author due to relative velocity structure of the $\lambda$2796 and $\lambda$2803 features.}{This system is of particular interest as it falls in the redshift range 5.523 $<$ z $<$ 5.879 which corresponds to a large Ly$\alpha$ trough discussed in detail by \cite{BECKER2015B}. This is {one of two} possible metal absorption systems in the corresponding redshift range. However, there are no other absorbers associated with this \ion{Si}{IV} system and both transitions are heavily blended. Is this then a real system?}

{We find that this is the only \ion{Si}{IV} system with no other associated absorbers such as \ion{C}{IV}, \ion{Si}{II} or \ion{Fe}{II}. All other \ion{Si}{IV} systems identified in this work as well as those of B15\footnote{see their Tables 2-10} and D13 have at least one other associated absorber. Given this extra information, we suspect that this individual \ion{Si}{IV} absorber is a false positive. However, since we do not use such criteria for the selection of other absorbers and this system does meet our general selection criteria described in Sections \ref{sec:autosearch} and \ref{sec:visualcheck}, we include it in the absorber population statistics. These statistics are adjusted for {\it user success/failure, false positive contamination} and {\it completeness} as described in Section \ref{sec:surveycompandfp}. }

{The second system in the redshift range 5.523 $<$ z $<$ 5.879} is system 8, a single component absorber anchored by \ion{Si}{IV} with associated \ion{Si}{II}$\lambda$1260. We note that System 8 is a marginal detection as both transitions are affected by skyline residuals. The presence of the associated \ion{Si}{II} $\lambda$1260 lends credibility to the detection but, itself, was identified as a possible \ion{Mg}{II} doublet by the automatic detection algorithm. It was not selected by the lead author as the $\lambda\lambda$2796 2803 transitions were not well matched when considering the full absorption profile composed of the \ion{Si}{II} $\lambda$1260 component of System 8 and the \ion{Si}{IV} $\lambda$1393 component of System 6.

System 2 is anchored by \ion{C}{IV} (\ion{C}{IV} $\lambda$1548 blended with \ion{Al}{II} $\simeq$ 4.45996) and \ion{Mg}{II} doublets. We identify associated \ion{Fe}{II} (\ion{Fe}{II} $\lambda$2585 blended with a skyline), \ion{Mg}{I}, \ion{Si}{II} (blended with \ion{Fe}{II} at z $\simeq$ 2.47759) and \ion{Al}{II}. System 3 is a three component \ion{C}{IV} system. System 4 is a two component \ion{C}{IV} system. System 6 is a two component system with \ion{C}{IV} and \ion{Si}{IV} absorbers. System 7 is a \ion{C}{IV} absorber whose $\lambda$1548 feature is blended with the \ion{C}{IV} $\lambda$1550 feature of System 6. 

\begin{table*}
	\centering
	\caption{Absorption systems identified in ULAS J0148+0600 sightline. A system is defined as all components within 500 kms$^{-1}$ of the lowest redshift component. Each component is marked with a letter id and all associated transitions are identified in the Ion column. The table also lists $z$, $W_0$, $log$($N$) and $b$ which are the redshift, equivalent width, column density and doppler parameter for each component Voigt profile fit. The 5$\sigma$ recovery selection criteria are defined in eq. \ref{eq:barc}. No lower bound is presented for systems with $b$ = 1 kms$^{-1}$ as the minimum doppler parameter we allow for a Voigt Profile is 1 kms$^{-1}$.}	
	\label{tab:u0148table}

	\begin{tabular}{c c c c c c c c} 
		
		&	 & & \Large{ULAS J0148+0600} & 	& 					&			& 				 	\\ 
		\hline

		Sys & ID & Ion 				 & z 			& $W_0$ & log(\textrm{N/cm$^2$})			& b						& 	5$\sigma$	recovery \\ 
		
		& & &			& 	( \AA)		 & & ( kms$^{-1}$)		 & rate	\\ 
		
		\hline 
		
		1 & a & CIV 1548 & 4.57120 $\pm$ 9.0 $\times 10^{-5}$ & 0.042 $\pm$ 0.002 & $13.06_{-0.26}^{+ 0.45}$ & $25.0_{-6.77}^{+ 64.1}$ & 0.93 \\ 
		& & CIV 1550 & & 0.020 $\pm$ 0.001 & & 	 & \\

		2 & a & CIV 1548$\Downarrow$ & 4.89095 $\pm$ 0.00015 & 0.059 $\pm$ 0.002 & $13.24_{-0.16}^{+ 0.38}$ & $31.8_{-4.46}^{+ 67.6}$ & 0.96 \\ 
		& & CIV 1550 & & 0.031 $\pm$ 0.003 & & 	 & \\ 
		& & MgI 2852 & 4.89099 $\pm$ 6.8 $\times 10^{-5}$ & 0.121 $\pm$ 0.006 & $11.98_{-0.20}^{+ 0.27}$ & $47.7_{-8.96}^{+ 22.1}$ & \\ 
		& & SiII 1526$\Downarrow$ & 4.89049 $\pm$ 3.7 $\times 10^{-5}$ & 0.233 $\pm$ 0.002 & $14.43_{-0.08}^{+ 0.15}$ & $17.8_{-1.24}^{+ 0.48}$ & \\ 
		& & AlII 1670 & 4.89054 $\pm$ 0.00013 & 0.268 $\pm$ 0.004 & $13.05_{-0.06}^{+ 0.06}$ & $27.3_{-1.92}^{+ 1.42}$ & \\ 
		& & FeII 2600 & 4.89056 $\pm$ 7.5 $\times 10^{-5}$ & 0.522 $\pm$ 0.004 & $14.07_{-0.08}^{+ 0.09}$ & $23.7_{-1.82}^{+ 3.84}$ & \\ 
		& & FeII 2586$\Uparrow$ & & 0.307 $\pm$ 0.008 & & 	 & \\ 
		& & FeII 2382$\Uparrow$ & & 0.551 $\pm$ 0.011 & & 	 & \\ 
		& & FeII 1608 & & 0.110 $\pm$ 0.003 & & 	 & \\ 
		
		& b & SiII 1526 & 4.89172 $\pm$ 4.3 $\times 10^{-5}$ & 0.223 $\pm$ 0.002 & $14.30_{-0.06}^{+ 0.07}$ & $20.0_{-1.54}^{+ 0.67}$ & \\ 
		& & AlII 1670 & 4.89175 $\pm$ 0.00012 & 0.243 $\pm$ 0.005 & $13.02_{-0.06}^{+ 0.06}$ & $22.5_{-2.13}^{+ 1.53}$ & \\ 
		& & FeII 2600 & 4.89176 $\pm$ 5.7 $\times 10^{-5}$ & 0.208 $\pm$ 0.003 & $14.30_{-0.08}^{+ 0.53}$ & $7.04_{-0.17}^{+ 1.65}$ & \\ 
		& & FeII 2586 & & 0.181 $\pm$ 0.044 & & 	 & \\ 
		& & FeII 2382 & & 0.223 $\pm$ 0.010 & & 	 & \\ 
		& & FeII 1608 & & 0.096 $\pm$ 0.003 & & 	 & \\ 
		3 & a & CIV 1548 & 4.93212 $\pm$ 0.00011 & 0.093 $\pm$ 0.002 & $13.48_{-0.10}^{+ 0.00}$ & $18.7_{-3.79}^{+ 5.74}$ & 0.97 \\ 
		& & CIV 1550 & & 0.052 $\pm$ 0.002 & & 	 & \\ 
		& b & CIV 1548 & 4.93308 $\pm$ 0.00020 & 0.057 $\pm$ 0.002 & $13.22_{-0.16}^{+ 0.07}$ & $20.8_{-3.77}^{+ 9.50}$ & 0.96 \\ 
		& & CIV 1550 & & 0.029 $\pm$ 0.002 & & 	 & \\ 
		& c & CIV 1548 & 4.93463 $\pm$ 0.00030 & 0.042 $\pm$ 0.003 & $13.07_{-0.27}^{+ 0.25}$ & $38.3_{-10.1}^{+ 8.97}$ & 0.95 \\ 
		& & CIV 1550 & & 0.022 $\pm$ 0.003 & & 	 & \\

		4 & a & CIV 1548 & 4.95183 $\pm$ 0.00033 & 0.013 $\pm$ 0.002 & $12.56_{-0.69}^{+ 0.73}$ & $19.2_{-18.2}^{+ 66.5}$ & 0.71 \\ 
		& & CIV 1550 & & 0.006 $\pm$ 0.002 & & 	 & \\ 
		& b & CIV 1548 & 4.95809 $\pm$ 0.00027 & 0.012 $\pm$ 0.002 & $12.56_{-0.42}^{+ 0.68}$ & $3.58_{-2.48}^{+ 77.6}$ & 0.71 \\ 
		& & CIV 1550 & & 0.006 $\pm$ 0.002 & & 	 & \\ 
		
		5 & a & CIV 1548 & 5.02327 $\pm$ 0.00013 & 0.092 $\pm$ 0.010 & $13.49_{-0.11}^{+ 0.13}$ & $16.1_{-5.47}^{+ 7.78}$ & 0.93 \\ 
		& & CIV 1550 & & 0.052 $\pm$ 0.008 & & 	 & \\ 
		
		6 & a & CIV 1548 & 5.12488 $\pm$ 6.9 $\times 10^{-5}$ & 0.199 $\pm$ 0.005 & $13.87_{-0.07}^{+ 0.07}$ & $28.0_{-2.04}^{+ 0.61}$ & 0.97 \\ 
		& & CIV 1550$\Downarrow$ & & 0.119 $\pm$ 0.004 & & 	 & \\ 
		& & SiIV 1393 & 5.12509 $\pm$ 6.7 $\times 10^{-5}$ & 0.040 $\pm$ 0.001 & $12.75_{-0.21}^{+ 0.26}$ & $24.3_{-5.55}^{+ 15.7}$ & 0.96 \\ 
		& & SiIV 1402 & & 0.022 $\pm$ 0.001 & & 	 & \\ 
		
		7 & a & CIV 1548$\Downarrow$ & 5.13495 $\pm$ 0.00024 & 0.076 $\pm$ 0.005 & $13.33_{-0.18}^{+ 0.20}$ & $35.1_{-8.21}^{+ 30.2}$ & 0.94 \\ 
		& & CIV 1550$\Downarrow$ & & 0.036 $\pm$ 0.003 & & 	 & \\

		8 & a & SiIV 1393 & 5.77495 $\pm$ 0.00038 & 0.029 $\pm$ 0.013 & $12.63_{-0.22}^{+ 0.21}$ & $6.54_{-5.42}^{+ 7.25}$ & 0.64 \\ 
		& & SiIV 1402 & & 0.016 $\pm$ 0.005 & & 	 & \\ 
		& & SiII 1260 & 5.77515 $\pm$ 8.5 $\times 10^{-5}$ & 0.003 $\pm$ 0.006 & $12.23_{-0.20}^{+ 0.20}$ & $13.7_{-7.97}^{+ 6.36}$ & \\

		9 & a & SiIV 1393$\Downarrow$ & 5.82630 $\pm$ 0.00013 & 0.035 $\pm$ 0.002 & $12.71_{-0.12}^{+ 0.14}$ & $9.70_{-6.98}^{+ 3.12}$ & 0.68 \\ 
		& & SiIV 1402$\Downarrow$ & & 0.019 $\pm$ 0.001 & & 	 & \\ 
		\hline 
	\end{tabular}

	\begin{flushleft}

		$\Downarrow$ $\hspace{0.8mm}$ denotes a component with a blended feature
		
		$\Uparrow$ $\hspace{0.8mm}$ denotes a component polluted by a sky-line or poor subtraction residual
		
		$^{\ast}$ $\hspace{1.2mm}$ denotes system which does not meet our 5$\sigma$ recovery selection criteria
	\end{flushleft}
\end{table*} 


\subsubsection{SDSS J0927+2001}
\label{sec:s0927} 

We present 10 new systems and the highest redshift absorber in this sightline which meets our 5$\sigma$ recovery selection criteria is system 10 with z = 5.66382 $\pm$ 0.00020. {All systems and associated components are presented in detail in Section B of the online Appendix.}

Systems 1, 2$^{\ast}$, 3, 6, 7 and 9 are single component \ion{C}{IV} systems. Systems 4 and 10 are two component \ion{C}{IV} systems. System 5 is a 3 component \ion{C}{IV} system where the third component has associated \ion{Al}{II}$\lambda$1670 which is heavily blended with a \ion{Mg}{II} system with {z = 2.34879 $\pm$ 1$\times$10$^{-5}$}. System 5 also has an associated \ion{Mg}{II} absorber. System 8 is a single component system with both \ion{C}{IV} and \ion{Si}{IV} absorbers.

\subsubsection{SDSS J1306+0356}
\label{sec:s1306} 

This sightline was previously investigated by D13 who confirm systems 2, 4, 8 and 9 first identified by \cite{SIMCOE2011} (S11). D13 first discover systems 1, 3, 5 and 6. The highest redshift absorber in this sightline which meets our 5$\sigma$ recovery selection criteria is system 15 with z = 5.80738 $\pm$ 0.00017. {All systems and associated components are presented in detail in Section C of the online Appendix.}

We present seven new systems 7, 10, 11$^{\ast}$, 12, 13$^{\ast}$, 14$^{\ast}$ and 15 with four of them passing our 5$\sigma$ recovery selection criteria. Systems 5, 6, 7, 10, 11$^{\ast}$, 12, 13$^{\ast}$, 14$^{\ast}$ and 15 are single component \ion{C}{IV} systems. Systems 1 (\ion{C}{IV} $\lambda$1550 blended with the \ion{Si}{II}$\lambda$1526 feature of system 2), 3 and 4 are two component \ion{C}{IV} systems.

System 2 is a complex absorber anchored by \ion{C}{IV} and \ion{Mg}{II} doublets with associated \ion{Al}{II}, \ion{Fe}{II} and \ion{Si}{II}. {We note that the \ion{Al}{II}$\lambda$1670 component $b$ is most likely blended with an unidentified transition which results in an artificial broadened profile. This results from the fact that we do not tie the Doppler $b$ parameter of the ionisation systems during the Voigt profile fitting}. 

S11 suggest a possible \ion{C}{IV} system at z = 4.702 and it is flagged by the automatic detection algorithm (Sec. \ref{sec:autosearch}) but the $\lambda$1548 feature is contaminated by a sky-line while the possible $\lambda$1550 feature is actually the $\lambda$1548 feature of system 5. For these reasons, this candidate is rejected by the lead author in the visual inspection step (Sec. \ref{sec:visualcheck}). During the same visual inspection step, the lead author also does not select system 2 presented by D13 (z = 4.58040 $\pm$ 7$\times$10$^{-5}$) as the $\lambda$1550 feature is not well matched with the $\lambda$1548's velocity structure.

System 8 is a complex absorber anchored by \ion{C}{IV} and \ion{Mg}{II} doublets with associated \ion{Al}{II}, \ion{Mg}{I}, \ion{Fe}{II} and \ion{Si}{II} as first suggested by S11. System 9 is a similarly complex absorber with much weaker \ion{Mg}{I} which we were not able to fit confidently. The systems could not be fit when the continuum was adjusted by +0.05. These two systems were discussed in C17 as they prove challenging to the system definition (all components within 500 kms$^{-1}$) and system 9 is omitted by \citet{CHEN2016} from their analysis. In order to highlight these systems, we plot them together in Figure \ref{fig:S1306_z_4d85878}. We anchor the velocity scale on the bluest \ion{C}{IV} component (z = 4.85878 $\pm$ 3.5$\times$10$^{-5}$) and then plot vertical dashed lines at three 500 kms$^{-1}$ intervals. As can be seen, the reddest component of system 8 are within 500 kms$^{-1}$ of the bluest component of system 9 but each system has distinct velocity structures. {We also plot vertical dashed lines at +600 and +900 kms$^{-1}$ and we discuss these systems in more detail in Section \ref{sec:sys8and9discussion}. }

\subsubsection{ULAS J1319+0959}
\label{sec:u1319} 

This sightline was previously investigated by D13 who confirm systems 6 and 9 first identified by S11 and first discover systems 1, 2, 3, 4 and 8. {D13 treat our system 2 as two systems given that their distinct components are separated by $\sim$350 kms$^{-1}$. However, in order to be consistent with our definition of a system\footnote{all components within 500 kms$^{-1}$} we treat this as a single system.} The highest redshift absorber in this sightline which meets our 5$\sigma$ recovery selection criteria is system 9 with z = 5.57037 $\pm$ 0.00033. {All systems and associated components are presented in detail in Section D of the online Appendix.}

We present two new systems, 5 and 7$^{\ast}$. We only use system 5 in our analysis as system 7 does not meet our 5$\sigma$ recovery selection criteria. Systems 1, 2, 4, 5 and 7$^{\ast}$ are single component \ion{C}{IV} systems. System 2 (\ion{C}{IV} $\lambda$1550 of component $b$ is blended with the \ion{C}{IV} $\lambda$1548 components of system 3) is a two component \ion{C}{IV} system.

System 3 is anchored by both \ion{C}{IV} and \ion{Mg}{II}. The \ion{C}{IV} $\lambda$1548 features are blended with the \ion{C}{II} component of system 9 and the \ion{C}{IV} $\lambda$1550 of system 2. The \ion{C}{IV} $\lambda$1550 features are blended with the \ion{Si}{IV} $\lambda$1402 features of system 6. System 6 is a two component \ion{Si}{IV} system where component $b$ also has associated \ion{C}{IV}. System 8 is a two component system anchored by \ion{C}{IV}, \ion{Si}{IV} and \ion{Mg}{II}. Component $b$ also has associated \ion{Al}{II}$\lambda$1670.

System 9 is a three component system anchored by \ion{Si}{IV}. The \ion{C}{IV} associated with system 9 is not detected by the automatic detection algorithm at a 5$\sigma$ cutoff but component $b$ is detected at a 3$\sigma$ selection. We introduce the absorber as it was previously discovered by S11 and D13 and has associated \ion{Si}{IV} doublet to also anchor the redshift of the absorber. We discuss the impact of this decision on our resulting incidence line statistics (Sec. \ref{sec:incidencerates}), comoving mass density calculation (Sec. \ref{sec:omega}) and resulting column density distribution functions (Sec. \ref{sec:cddf}). We confirm the possible associated \ion{C}{II} first suggested by S11 but we do not identify any associated \ion{Fe}{II}. We do identify associated \ion{Al}{II}$\lambda$1670.

We are not able to confirm the D13 systems 2 (z = 4.62931 $\pm$ 8$\times$10$^{-5}$) and 6 (z = 4.70325 $\pm$ 2$\times$10$^{-5}$) as they are not flagged by our automatic detection algorithm and do not have other associated transitions to solidify the identification. 

\section{Survey completeness and false positive corrections}
\label{sec:surveycompandfp}

 When considering the statistics of absorption line systems we must account for the wavelength dependent signal-to-noise profile (visual vs. near-infrared) as well as the strength of the absorber (equivalent width or column density). Furthermore, one must account for the human impact if a visual inspection/user voting process selects from the output of an automatic detection algorithm. In summary, we use the prescriptions put forth in C17, which follow the steps described in \citet{MATEJECK2012}.

We first create a library of \ion{C}{IV} and \ion{Si}{IV} absorbers which are inserted at every \AA $ $ in the spectra of each QSO. We then search for them using the same automatic detection algorithm described in Sec. \ref{sec:autosearch}. Following this, the lead author quantifies their ability to accurately identify true absorbers ($user$ $success$) as well as incorrectly identify random features or artificially spaced doublets\footnote{described in Section \ref{sec:fps}} as true absorbers ($user$ $failure$). $User$ $success$ and $user$ $failure$ are computed as functions of S/N. We then turn these $user$ $success$ and $user$ $failure$ likelihoods into 2D maps with the same resolution as the recovery grids outputted by the automatic detection algorithm. We combine these and create recovery maps adjusted for $user$ $success/failure$ for each sightline for both \ion{C}{IV} and \ion{Si}{IV} doublets.

In order to account for false positive contamination, we repeat the entire analysis and completeness calculations for the artificially spaced doublets: \ion{C}{IV'} ($\lambda\lambda$1548.2049 1553.3519) and \ion{Si}{IV'} ($\lambda\lambda$1393.76018 1411.78580). We then combine the results of this search and analysis of artificially spaced doublets with that of the true \ion{C}{IV} and \ion{Si}{IV} absorbers to compute a single scalar which accounts for variable completeness across a redshift bin, the strength of the absorber, false positive contamination as well as the impact of the human interaction step (see $A$ in Table \ref{tab:incidenceomegatable}). We present the details of each step below. 

\subsection{Automatic recovery of doublets}
 \label{sec:autorecovery} 
 
In order to identify the true number of absorbers for a given redshift path, $dz$, we must first identify for what fraction of that path we can confidently identify the discovered absorbers. This recovery fraction is a function of both the strength of the absorber and the signal-to-noise profile of each quasar spectrum considered. As a first step, we use \textsc{RDGEN} \citep{RDGEN} to extract a library of Voigt profiles consistent with the observed components. For \ion{C}{IV}, we extract absorbers with a column density range 12.5 $\le$ log(N$_{\textrm{sys}}$/cm$^{2}$) $\le$ 14.5 in 0.1 increments. For \ion{Si}{IV}, we extract absorbers with a column density range 12.2 $\le$ log(N$_{\textrm{sys}}$/cm$^{2}$) $\le$ 13.8 in 0.1 increments. For each column density, we extract six profiles with a $b$ parameter value in the range 10 $\le$ b $\le$ 60 kms$^{-1}$ with a step size of 10 kms$^{-1}$. We sample every \AA $ $ where \ion{C}{IV} or \ion{Si}{IV} can be observed by injecting an absorption system in 100 \AA $ $ steps. In total, we insert and search for 444288 \ion{Si}{IV} and 1078200 \ion{C}{IV} absorbers.

 {The injected systems are searched for automatically using the same simple algorithm used to create the initial candidate lists described in sec \ref{sec:autosearch}. The output of the detection algorithm is a Heaviside function, $H$(log(N), $b$, $z$), where log(N$_{\textrm{sys}}$/cm$^{2}$), $b$ and $z$ are the column density, {Doppler} parameter and redshift location of the inserted \ion{C}{IV} and \ion{Si}{IV} Voigt profile. The values of the output are }

\begin{equation} 
H(\log(\textrm{N}, b, z) = \left\{
 \begin{array}{ll}
 0 &\textrm{if the injected system is not detected, } \\
 1 &\textrm{if the injected system is detected, }
 \end{array}
 \right.
	\label{eq:heaviside}
\end{equation}

\noindent Next, we bin the above Heaviside function across all $b$ values and redshift bins ($dz$ = 0.01) and compute the recovery fraction of an inserted doublet as 

\begin{equation}
 L(\log(\textrm{N}, dz_j) = \frac{1}{n} \sum\limits_{i = 1}^{n} H(\log(\textrm{N}), b, z)\
	\label{eq:lrecovery}
\end{equation}

\noindent where $n$ is the total number of inserted systems such that $z \in dz_j$ with column density log(N$_{\textrm{sys}}$/cm$^{2}$). This above expression denotes how often an inserted absorber would make it on to the initial candidate list which was then visually inspected by the lead author.

\subsection{Adjusting for user interaction}
 \label{sec:userinteraction} 
 
In order to account for the ability of the lead author to accurately identify an absorber, we create a simple simulation which randomly chooses to/not to insert an absorber with/without a true rest-frame separation. This randomisation ensures the that there is no a priori expectation on whether an absorber is inserted and if so, if it is a true absorber. We will describe the $false$ absorbers in the following section (\ref{sec:fps}). This voting process is {run by eye for 12000 instances}. The results (1-for discovery, 0-no discovery) are binned as a function of the boxcar S/N of the inserted feature SNR$\equiv W_{d1}/\sigma W_{d1}$\footnote{$d1$ denotes the $\lambda$1548 or $\lambda$1393 feature of the \ion{C}{IV} and \ion{Si}{IV} true and false absorbers.}.

Next, we fit the $user$ $success$ (using a $\chi^{2}$ minimisation technique) with an exponential function of the form

\begin{equation}
 P_{d1}(SNR) = P_{\infty} (1-e^{S/SNR})
	\label{eq:usersuccess}
\end{equation}
\begin{figure}
	\begin{center}
		\includegraphics[width=8.4cm]{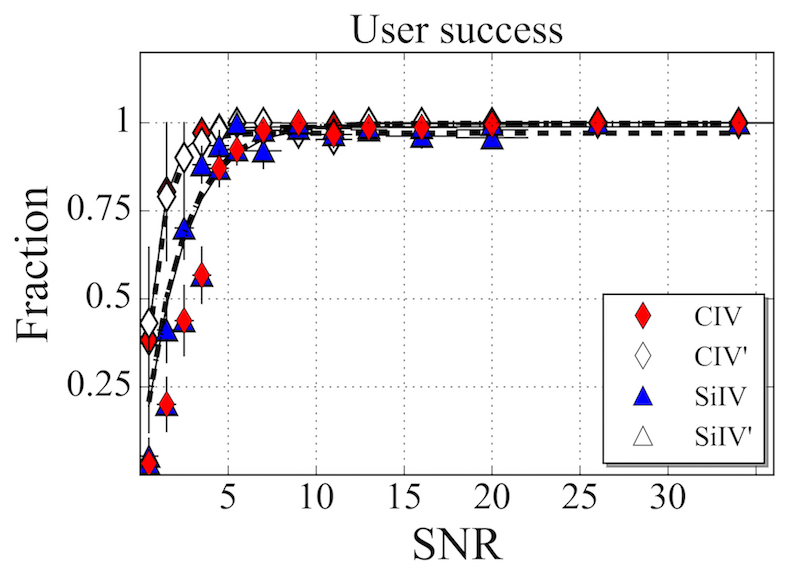} 
		\includegraphics[width=8.4cm]{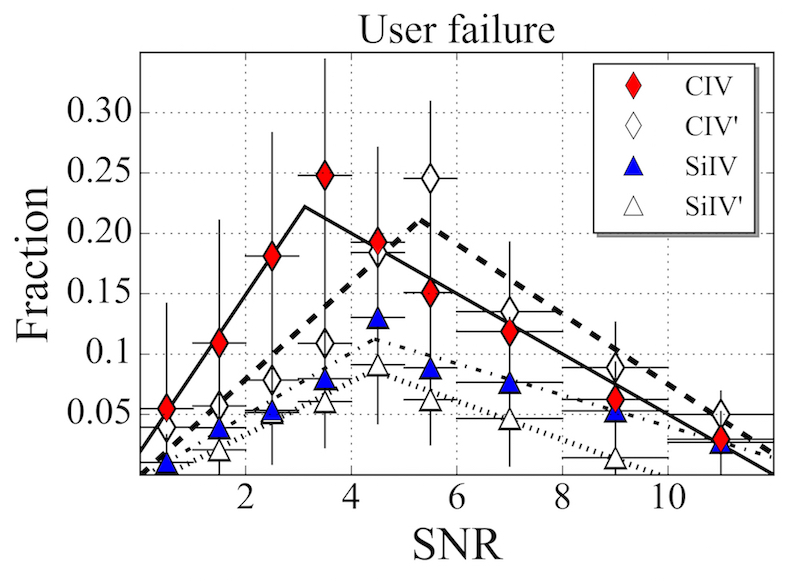} 

	 	\caption{The binned $user$ $success$ (top panel; eq. \ref{eq:usersuccess}) and user failure (bottom panel; eq. \ref{eq:userfailure}) for \ion{C}{IV}, \ion{C}{IV'}, \ion{Si}{IV} and \ion{Si}{IV'}. Plotted with solid, dashed, dot-dash and dash-dash lines are the respective best fits. The horizontal bounds denote the S/N bin considered and the vertical error bars correspond to the associated 95$\%$ Wilson confidence interval.}
 		\label{fig:usersuccess}
	\end{center}
\end{figure}

\noindent where P$_{\infty}$ is the probability that the user will accept a true absorption system and S is an SNR exponential scale factor. Just as we found in C17 and similar to the findings of \citet{MATEJECK2012} and \citet{CHEN2016}, we find that even for the best S/N regions, the user acceptance rate is not 100$\%$, except for the artificial doublet \ion{Si}{IV'}.

Following this, we next fit the $user$ $failure$ with a triangle function

\begin{equation}
 P^{FP}(SNR) = \left\{
 \begin{array}{ll}
 P^{FP}_{max} \left(SNR/s_{p}\right) & SNR \le s_{p} ,\\
 P^{FP}_{max} \left(\frac{SNR-s_{f}}{s_{p}-s_{f}}\right) & SNR>s_{p} ,
 \end{array}
 \right.
	\label{eq:userfailure}
\end{equation}

\noindent where $P^{FP}_{max}$ is the maximum contamination rate which arises at $s_{p}$. We find that the user acceptance of injected false doublets as real approaches 0 as the SNR reaches $\sim$13. The binned values and best fits can be seen in Figure \ref{fig:usersuccess}. All best fit values can be seen in Table \ref{tab:usersuccessfailure}.

\begin{table}
 \centering
 \caption{$User$ $success$ (eq. \ref{eq:usersuccess}) and $user$ $failure$ (eq. \ref{eq:userfailure}) best fit parameters. The best fits and binned values can be seen in Figure \ref{fig:usersuccess}.}
 \label{tab:usersuccessfailure}
 \begin{tabular}{|| c | c | c | c | c ||}
 \hline
 ion & P$_{\infty}$ & $S$ & $P^{FP}_{max}$ & $s_p$\\

 \hline
 \ion{C}{IV}& 0.97 & 0.99 & 0.22 & 3.12 \\
 \ion{C}{IV'} & 0.98 & 0.93 & 0.21 & 5.30 \\ 
 \ion{Si}{IV}& 0.99 & 2.16 & 0.08 & 4.42 \\
 \ion{Si}{IV'} & 1.00 & 2.30 & 0.11 & 4.43 \\

 \end{tabular}
\end{table}

Finally, we turn the $user$ $success$ and $user$ $failure$ functional values (eqs. \ref{eq:usersuccess} and \ref{eq:userfailure}) into grids binned with the same resolution as the recovery function (eq. \ref{eq:lrecovery}). The resulting grids are denoted as $A(log(\textrm{N}), dz_j)$ and $A^{FP}(log(\textrm{N}), dz_j)$, respectively. Following this, we combine the recovery rate (from the automatic detection) with the $user$ $success$ grids

\begin{equation}
 C(\log(\textrm{N}), dz_j) = L(\log(\textrm{N}), dz_j) \times A_{d1}(\log(\textrm{N}), dz_j)\
	\label{eq:recoveryhuman}
\end{equation}

\begin{figure}
	\includegraphics[width=8.4cm]{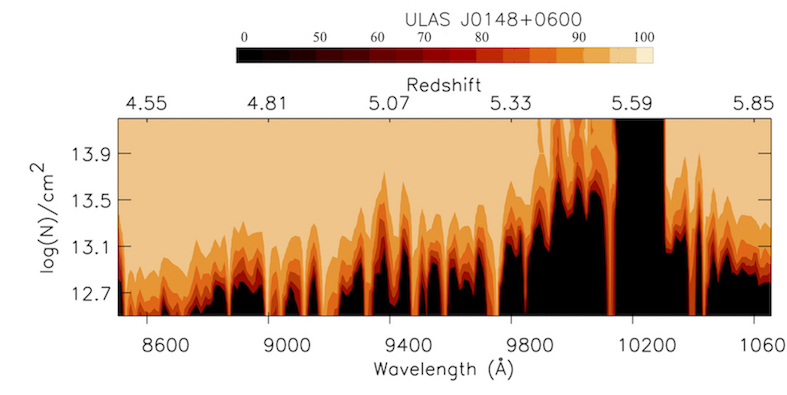}
	\caption{Example \ion{C}{IV} 5$\sigma$ completeness test results for the ULAS J0148+0600 sightline. The $x$ and $y$ axis represent the wavelength ( \AA) and column density (log(\textrm{N})/cm$^{2}$), respectively, of inserted systems. The top $x$ axis represents the corresponding redshift of an inserted system. Their recovery rate, C($dN_{1548}$, $dz$), is denoted by the color bar plotted in each panel. All recovery rates below 50$\%$ are shaded in black. {Recovery rates for all sightlines and for both {\it real} and {\it artificial} doublets are presented in Sections F and G of the online appendix}. The wavelength resolution of the recovery function, C($dN_{1548}$, $dz$), allows us to identify clean portions of the spectra as can be seen in panel $A$ at around $\sim$10100 \AA $ $ $ $ where the recovery rate is above 50$\%$ for even the weakest of inserted systems. } 
	
	\label{fig:c4comps5}
\end{figure}

\noindent and {an example} can be seen \ref{fig:c4comps5}. The recovery fraction corrected for $user$ $success$ associated with each doublet can be seen in the 5$\sigma$ recovery rate columns of each sightline discovery table available {in the online appendix (see Table \ref{tab:u0148table} for an example)}. Thirty seven \ion{C}{IV} systems and seven \ion{Si}{IV} systems survive a 5$\sigma$ cutoff. {For our analysis, we only use those systems with at least one component with 5$\sigma$ 50$\%$ or greater recovery fraction corrected for $user$ $success$, except for system 9 in $ULAS$ $J1319+0959$ for reasons described in sub-section \ref{sec:u1319}.}

 \subsection{False positives}
 \label{sec:fps}
In order to quantify the contamination by false positive doublets, we search for the artificially spaced absorbers: \ion{C}{IV'}($\lambda\lambda$1548 1553) and \ion{Si}{IV'}($\lambda\lambda$1393 1411). We perform this investigation in the same fashion as the search for the \ion{C}{IV} and \ion{Si}{IV} doublets. 

We first create an initial candidate list using the same detection algorithm described in Sec. \ref{sec:autosearch}. The automatic detection algorithm outputs 76 \ion{C}{IV'} and 32 \ion{Si}{IV'} candidates. The lead author then selects 3 \ion{C}{IV'} and 1 \ion{Si}{IV'} possible absorbers. {These absorbers are selected for their similar velocity profile, with the same considerations described in sub-section \ref{sec:visualcheck}.} Following this, we adjust \textsc{VPFIT} to fit the above artificial doublets and measure their column density and Doppler parameter. 

Next, we extract a library of artificial ions which we then insert and search for in the same manner and with the same resolution as described in sec. \ref{sec:autorecovery}. The results are then binned and adjusted for $user$ $success$ (see eq. \ref{eq:recoveryhuman}). The resulting maps can be seen {in the online appendix}. {As with the physically spaced doublets}, we only consider those artificial systems with a 5$\sigma$ recovery rate adjusted for $user$ $success$ $\ge$ 0.50. Two \ion{C}{IV'} and 1 \ion{Si}{IV'} artificial systems survive this selection. 

\subsection{Adjusting for varying completeness and false positives}
 \label{sec:compfps} Given the fine resolution of the recovery rate adjusted for $user$ $success$ (eq. \ref{eq:recoveryhuman} and Figure \ref{fig:c4comps5}) we must account for this variability across a larger bin of interest in order to derive meaningful statistics. As in C17, we first define a visibility function $R(dz_j)$. {It is defined as 1 redward of the Ly$\alpha$ emission peak to within 3000 kms$^{-1}$ of the QSO emission redshift of the $ion$ in question and 0 everywhere else.} We combine eq. \ref{eq:recoveryhuman} with this visibility function and re-define the recovery rate adjusted for $user$ $success$ 

 \begin{equation} 
 C(\log(\textrm{N}), dz_j) = \left\{
 \begin{array}{ll}
 0 &\textrm{if $C(\log(\textrm{N}), dz_j)$<0.50}, \\
 1 &\textrm{if $C(\log(\textrm{N}), dz_j)$ $\ge$ 0.50 $\&$ $R(dz_j)$ = 1. }
 \end{array}
 \right.
	\label{eq:rwz}
\end{equation} \begin{figure}
	\includegraphics[width=8.4cm]{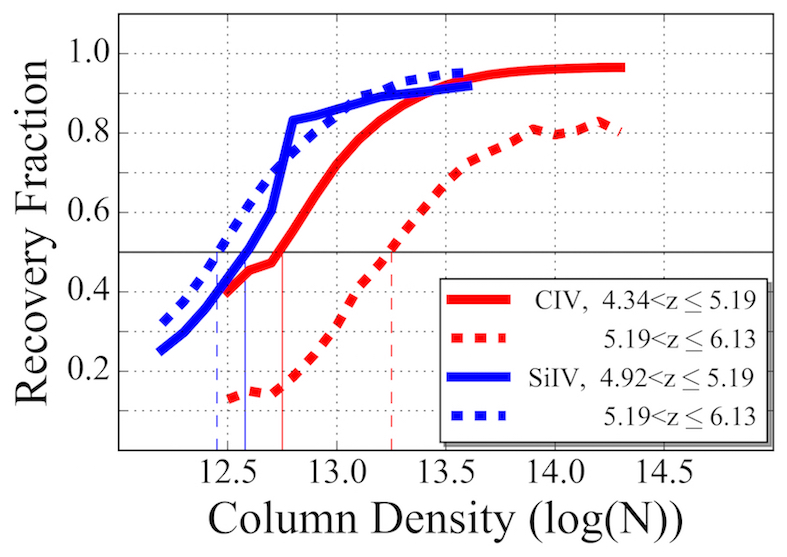}
 \caption{Recovery rates adjusted for $user$ $success$ (eq. \ref{eq:barc}) for \ion{C}{IV} and \ion{Si}{IV}. The vertical lines denote the log(N$_{\textrm{sys}}$/cm$^{2}$) values {down to} which we are 50$\%$ complete (12.45, 12.58, 12.75 and 13.25 respectively). The recovery of \ion{C}{IV} past z = 5.19 drops as the redshift pushes the absorbers into the near infra-red.}
 
 \label{fig:cbarbyzbin}
\end{figure}

 \noindent This formulation identifies all redshift bins ($dz_j$) in which a single component with column density log(N$_{\textrm{sys}}$/cm$^{2}$) will be identified in our analysis at least 50$\%$ of the time. Next, we split the \ion{C}{IV} redshift path in two redshift bins such that they cover a similar recovery adjusted redshift path. That redshift {midpoint} is at $z$ = 5.19. 
 
We then bin and find the average recovery rate adjusted for $user$ $success$ as a function of the column density log(N$_{\textrm{sys}}$/cm$^{2}$) of the inserted doublets \ion{C}{IV} and \ion{Si}{IV} (see Figure \ref{fig:cbarbyzbin}). For the respective redshift bins (below and above redshift 5.19), we find that for \ion{C}{IV}, we are at least 50$\%$ complete {down to} log(N$_{\textrm{sys}}$/cm$^{2}$) values of 12.75 and 13.25. For \ion{Si}{IV} we are at least 50$\%$ complete for log(N$_{\textrm{sys}}$/cm$^{2}$) values of 12.45 and 12.58 in the same redshift bins.

Following this, we define the average of a function in each redshift bin \begin{equation}
 	\bar{f}(dz_j) = \frac{\int \int R(dz_j) \times f(log(\textrm{N}), dz_j) \frac{d^2 N}{dzdlog\textrm{N}} dz dlog\textrm{N}}{\int \int R(dz_j) \frac{d^2 N}{dzdlog\textrm{N}} dz dlog\textrm{N}}\\
	\label{eq:barc}
\end{equation}

\noindent and compute the average recovery rate adjusted for $user$ $success$ ($\bar{C}$), the average recovery rate ($\bar{L}$), the average $user$ $success$ ($\bar{A}$) and average $user$ $failure$ ($\bar{A}^{FP}$). Using these values, we can then compute the true number of absorbers ($N$)

 \begin{equation}
 	N \equiv \frac{\ddot{N}} {\bar{C}-\bar{L} \times \bar{A}^{FP}} \\
	\label{eq:finalNcomp}
\end{equation}

\noindent In the same redshift bin, the true number of false positives ($N^{'}$) is \begin{equation}
 	N^{'} \equiv \frac{\ddot{N'}} {\bar{C'}-\bar{L'} \times \bar{A'}^{FP}} \\ 
	\label{eq:totalfalse}
\end{equation} 

\noindent thus, the false positive contamination rate is 

\begin{equation}
 	F = 1-\frac{N^{'}} {N} \\
	\label{eq:fprate}
\end{equation}

\noindent We combine eqs. \ref{eq:finalNcomp} and \ref{eq:fprate} and define a single scalar for each doublet which accounts for the variable completeness and false positive detections in a redshift bin ($A$; see Table \ref{tab:incidenceomegatable})

 \begin{equation}
 A = \frac{\bar{F}}{\bar{C}-\bar{L} \times \bar{A}^{FP}} \\
	\label{eq:abc}
\end{equation}

 {Finally, the true number of absorbers adjusted for completeness and false positive contamination is}

\begin{equation}
 	N = A \times \ddot{N} \\
	\label{eq:finalN}
\end{equation}

\noindent {with associated Poisson error}

\begin{equation}
 	\sigma N \equiv A \times \sqrt{\ddot{N}} \\
	\label{eq:finalNerror}
\end{equation}


\section{Absorption line statistics}
\label{sec:absstats}

When {investigating} the evolution of absorption systems observed in the spectra of QSOs, it is common to {calculate} their incidence rate ($dN/dz$), their comoving mass density ($\Omega_{ion}$) and their {column density distribution function }(CDDF). The incidence rate provide a simple accounting on the number of systems discovered over the total redshift path of a survey and the associated $\Omega_{ion}$ (eq. \ref{eq:omegaionfinal}) provides a direct measurement of all observed number of {atomic ions} over the same path (eq. \ref{eq:deltaxdistance}) normalised to the critical density today. Secondly, $\Omega_{ion}$ can also be measured by directly integrating the first moment of the associated CDDF (eq. \ref{eq:omegaintegral}). This method has the added benefit of investigating the impact of integration limits on the $\Omega_{ion}$ values. All incidence rates and associated $\Omega_{ion}$ values can be seen in Table \ref{tab:incidenceomegatable}. All CDDF best fit parameters and integration limits can be seen in Table \ref{tab:cddftable}.

\subsection{Incidence rates}
\label{sec:incidencerates} 

\begin{figure}
	\includegraphics[width=8.4cm]{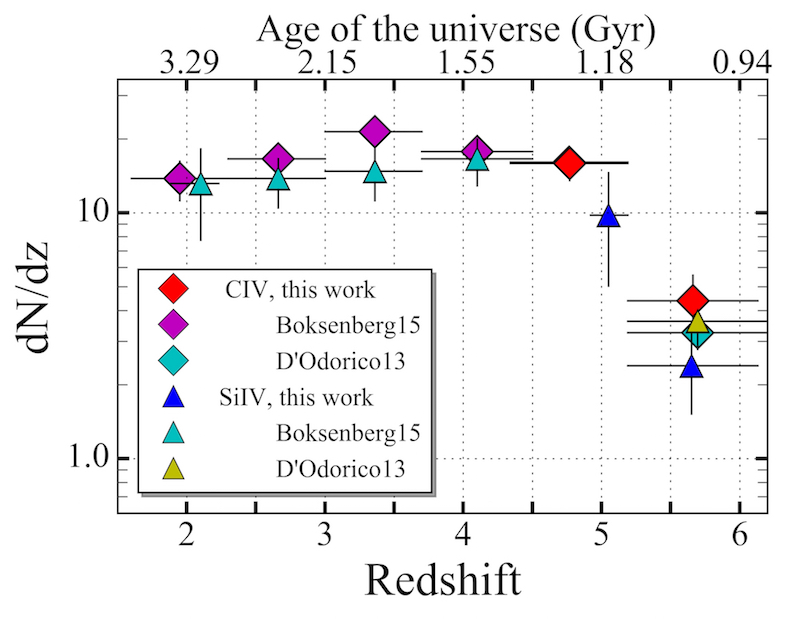}
	\caption{Incidence rates ($dN/dz$) for \ion{C}{IV} and \ion{Si}{IV}. The marker style and colors are described in the legend. All values measured with systems described in this work are presented in Table \ref{tab:incidenceomegatable}. The computation of the D13 values is described in the text. The B15 and S05 values have been adjusted to the {\it Planck} cosmology used in this work.}
	
 \label{fig:dndzdX}
\end{figure}

The true number of absorbers and associated error are computed in eqs. \ref{eq:finalN} and \ref{eq:finalNerror}. This accounts for the contamination by false positives and variable completeness across a redshift bin. The redshift bins are selected so that the \ion{C}{IV} path is split in almost half as described in sec. \ref{sec:compfps}. The incidence rate in a redshift bin ($dz$) is computed as

\begin{equation}
 	\Bigg(\frac{dN}{dz}\Bigg)=\frac{N\pm \sigma N}{dz}\\
	\label{eq:dndz}
\end{equation}

\noindent where $N$ is the completeness adjusted and false contamination corrected number of absorbers (eq. \ref{eq:finalN}) with associated error $\sigma N$ (eq. \ref{eq:finalNerror}). All incidence rate values computed in this work are presented in Table \ref{tab:incidenceomegatable} and can be seen Figure \ref{fig:dndzdX}.

For \ion{C}{IV}, we compute $dN/dz$ = 15.4 $\pm$ 2.8 at a median redshift <z> = 4.77 and find that it drops by almost a factor of $\sim$3.5 with $dN/dz$ = 4.4 $\pm$ 1.2 at a median redshift <z> = 5.66. For \ion{Si}{IV}, we find a similar evolution with the incidence rate $dN/dz$ = 9.8 $\pm$ 4.7 at a median redshift <z> = 5.05 dropping to $dN/dz$ = 2.4 $\pm$ 0.9 at a median redshift <z> = 5.66.

In order to compare with the survey data of D13 we search their tables A1, A2, A3, A4, A5 and A6. We count a total of 79 \ion{C}{IV} systems, 62 below redshift 5.19 ($dz$ = 3.86) and 17 above redshift 5.19 ($dz$ = 5.24). This leads to the \ion{C}{IV} incidence rates $dN/dz$ = 16.1 $\pm$ 2.0 at a mean redshift $z$ = 4.76 and $dN/dz$ = 3.2 $\pm$ 0.8 at a mean redshift $z$ = 5.69. They are in good agreement with our incidence rate (including system 9 in $ULAS$ $J1319+0959$). If system 9 is excluded, as it is the only manually introduced absorber as discussed in Sec. \ref{sec:u1319}, then $dN/dz$ = 3.7 $\pm$ 1.1. This value is still in good agreement with the D13 values.

We also count 19 \ion{Si}{IV} systems, all above redshift 5.19 across a redshift path $dz$ = 5.24. This leads to an incidence rate $dN/dz$ = 3.6 $\pm$ 0.8 at a mean redshift of $z$ = 5.69. Our incidence rates are well within the error bounds. {This is not surprising given that 2 out of our 4 sight lines are in common with D13, whose total sample size is 6 QSOs}. The incidence rates computed from the D13 tables are plotted in Figure \ref{fig:dndzdX}.

 Next, we consider if our definition of a system introduces a systematic difference between our incidence statistics and those of D13. We count system 5 in sightline $ULAS$ $J0148+0600$, system 10 in sightline $SDSS$ $J0927+2001$ and system 3 in sightline $SDSS$ $J1306+0356$ as individual systems {(see the online appendix)}. This definition could possibly underestimate the incidence rate in the respective redshift bins. However, this affects only 2 of the 37 \ion{C}{IV} systems below redshift 5.19 and only 1 of the 8 \ion{C}{IV} systems above redshift 5.19. The possible relative contribution to the incidence rates resulting from the system definition is then 5.4$\%$ and 12.5$\%$, respectively. We then find that our system definition does not introduce a systematic given that the respective relative Poisson errors (see Table \ref{tab:incidenceomegatable}) are several times larger.

Following this, we also compare with the incidence rates from B15. We find that both the incidence rates of \ion{C}{IV} and \ion{Si}{IV} exhibit a flat evolution until z$\sim$5. When we consider the incidence rates of B15 only, we compute a mean incidence rate for \ion{C}{IV} <$dN/dz$> = 17.4 $\pm$ 4.2. Similarly, for \ion{Si}{IV} we compute a mean incidence rate <$dN/dz$> = 14.6 $\pm$ 3.8. Both values are well within the error bounds of the incidence rates computed in this study below redshift 5.19. The B15 values are also plotted in Figure \ref{fig:dndzdX}.

Next, we compute the comoving incidence rates (\textit{dN/dX}) by calculating the absorption path of our survey, {noting that} a flat behavior in \textit{dN/dX} represents no comoving evolution in the cross-section of the population considered. The absorption distance is defined as 

\begin{equation}
X(z) = \frac{2}{3\Omega_{M}} \left[\Omega_{M}(1+z)^{3}+\Omega_{\Lambda}\right]^{1/2} \
	\label{eq:deltax}
\end{equation}

\noindent {thus for a redshift bin [$z_1$, $z_2$] with $z_2$>$z_1$, the absorption path between $z_2$ and $z_1$ is }

\begin{equation}
dX_{(z_1, z_2)} = X(z2)- X(z1)\
	\label{eq:deltaxdistance}
\end{equation}

For \ion{C}{IV} we compute $dN/dX$ = 3.6 $\pm$ 0.6 at a median redshift <z> = 4.77 and $dN/dX$ = 0.9 $\pm$ 0.3 at a median redshift <z> = 5.66 while for \ion{Si}{IV} we compute $dN/dX$ = 2.2 $\pm$ 1.1 at a median redshift <z> = 5.05 and $dN/dX$ = 0.5 $\pm$ 0.2 at a median redshift <z> = 5.66. We find that the comoving incidence rates of \ion{C}{IV} and \ion{Si}{IV} are not consistent with a no evolution scenario and both decrease from redshift $\sim$5 to 6.

\subsection{Comoving mass densities}
\label{sec:omega} 

The comoving mass density is defined as the first moment of the CDDF normalised to the critical density today 

\begin{equation}
 \Omega_{ion} = \frac{H_o m_{ion}} {c\rho_{crit}} \int Nf(N)dN \
	\label{eq:omegaintegral}
\end{equation}

\noindent where $m_{ion}$ is the mass of an ion, $\rho_{crit}$ = 1.89$\times$10$^{-29}\textrm{h}^2\textrm{g cm}^{-3}$ and $f(N)$ is the CDDF. In practice, it is approximated as 

\begin{equation}
\Omega_{ion} = \frac{H_o m_{ion}} {c\rho_{crit}} \frac{1}{dX} \sum_{s} \sum_{\ddot{N}} \sum_{k} N(ion) \
	\label{eq:omegaion}
\end{equation}

\noindent where $s$ represents the number of sightlines with $\ddot{N}$ discovered systems over absorption path $dX$, each with $k$ components with respective column density $N(ion)$. {Given that eq. \ref{eq:omegaion} accounts for all $k$ components of all $\ddot{N}$ discovered systems, it is not then susceptible to any bias resulting from the definition of a system.} In order to implement the same completeness and false positive corrections described in sub-section \ref{sec:compfps}, we combine eq. \ref{eq:omegaion} with eq. \ref{eq:abc} and calculate $\Omega_{ion}$ as

\begin{equation}
\Omega_{ion} = \frac{A}{dX} \frac{H_o m_{ion}} {c\rho_{crit}} \sum_{s} \sum_{\ddot{N}} \sum_{k} N_{sys}(ion) \
	\label{eq:omegaionfinal}
\end{equation}

All $A$, $dX$ and $\ddot{N}$ values are presented in Table \ref{tab:incidenceomegatable}. The associated errors are calculated by first bootstrapping 1000 times across the log(N$_{\textrm{sys}}$/cm$^{2}$) values of each system used. A system can be selected multiple times or not at all. The resulting errors correspond to the 66$\%$ confidence interval of the resulting distribution. The B15 and D13 values have been adjusted to the {\it Planck} cosmology used in this work.

\subsubsection{The comoving mass density of CIV}
\label{sec:omegac4}

\begin{figure}
	\includegraphics[width=8.4cm]{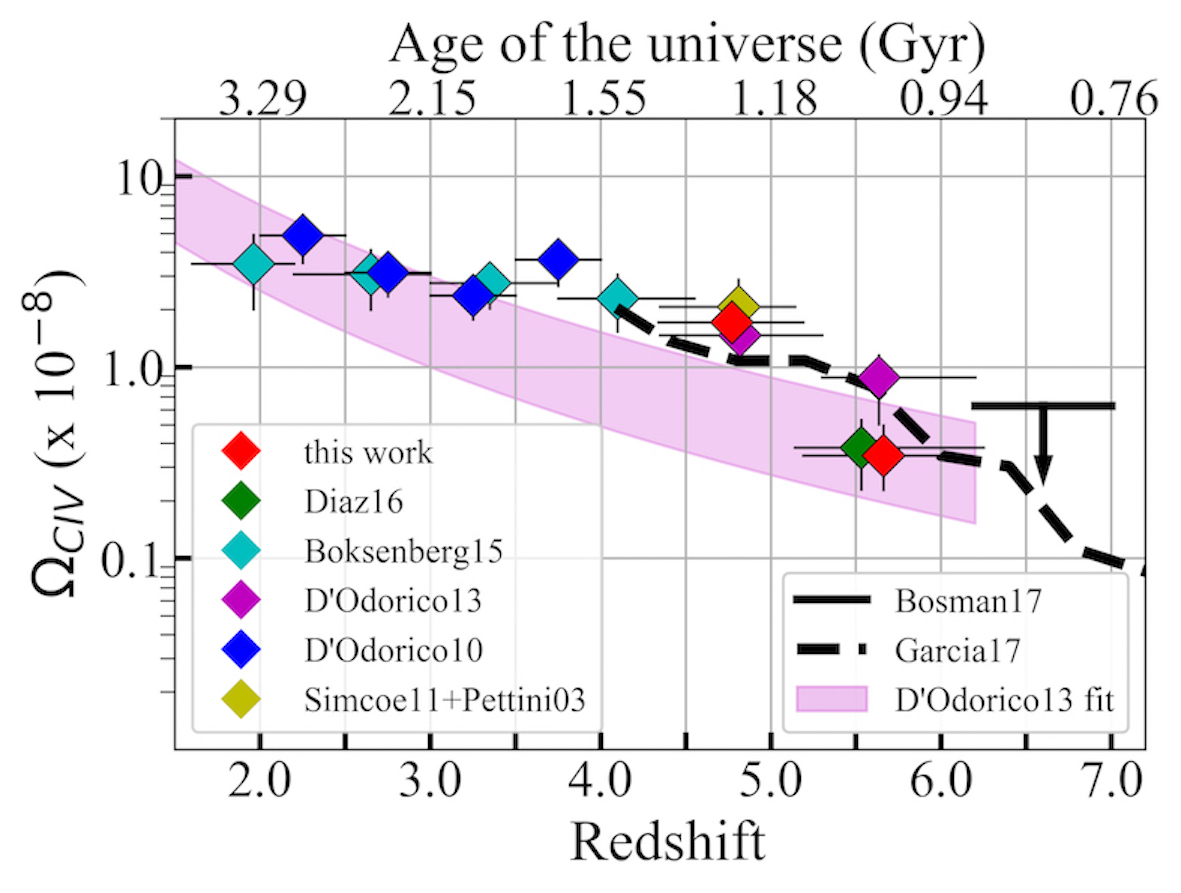}	
	  \caption{The evolution of the comoving mass densities of \ion{C}{IV} as measured in this work along with other observational values from literature and the simulation of G17. The line and marker style and colors are described in the legend.}

 \label{fig:omegac4values}
\end{figure}

We compute the comoving mass density of \ion{C}{IV} at the median redshifts <z> = 4.77 and <z> = 5.66 and find that, as in previous studies, it drops by approximately a factor of $\sim$4.7 from 1.6$^{+0.4}_{-0.1}$ $\times$10$^{-8}$ to 3.4$^{+1.6}_{-1.1}$ $\times$10$^{-9}$, respectively.

We plot our values along with those from other studies in Figure \ref{fig:omegac4values}. If we do not include system 9 in sightline $ULAS$ $J1319+0959$ we compute $\Omega_{\ion{C}{IV}}$ = 1.3$^{+0.4}_{-0.1}\times10^{-9}$. The {higher redshift} \citet{BOSMAN2017} value is an upper {limit} on $\Omega_{\ion{C}{IV}}$ at <z> = 6.6. {The dashed line represents the evolution of $\Omega_{\ion{C}{IV}}$ from the reference simulation of \cite{GARCIA2017}\footnote{Ch 18 512 MDW in their Table 1} (G17). In that work, a thousand lines of sight have been generated randomly inside the simulated box. The resulting spectra are first convolved with the instrumental resolution of the VLT-UVES spectrograph and then noise is added to reproduce the typical S/N of observational data. Within each sightline, \ion{C}{IV} individual absorption features have been searched for and fitted with \textsc{VPFIT}. 

The methodology described in G17 thus mimics the work flow of observational studies but is not impacted by observational constraints such as sky-line emission in the NIR. The synthetic absorption systems with column densities span the column density range 13.8 $\le$ log(N/cm$^{2}$) $\le$ 15.0 are summed up to calculate $\Omega_{\ion{C}{IV}}$. This column density range selection for synthetic absorbers follows the findings by D13. The \ion{C}{IV} systems discovered in this study spans 13.00 $\le$ log(N/cm$^{2}$) $\le$ 14.0.}

\subsubsection{The comoving mass density of SiIV}
\label{sec:omegasi4}
\begin{figure*}
	\includegraphics[width=16cm]{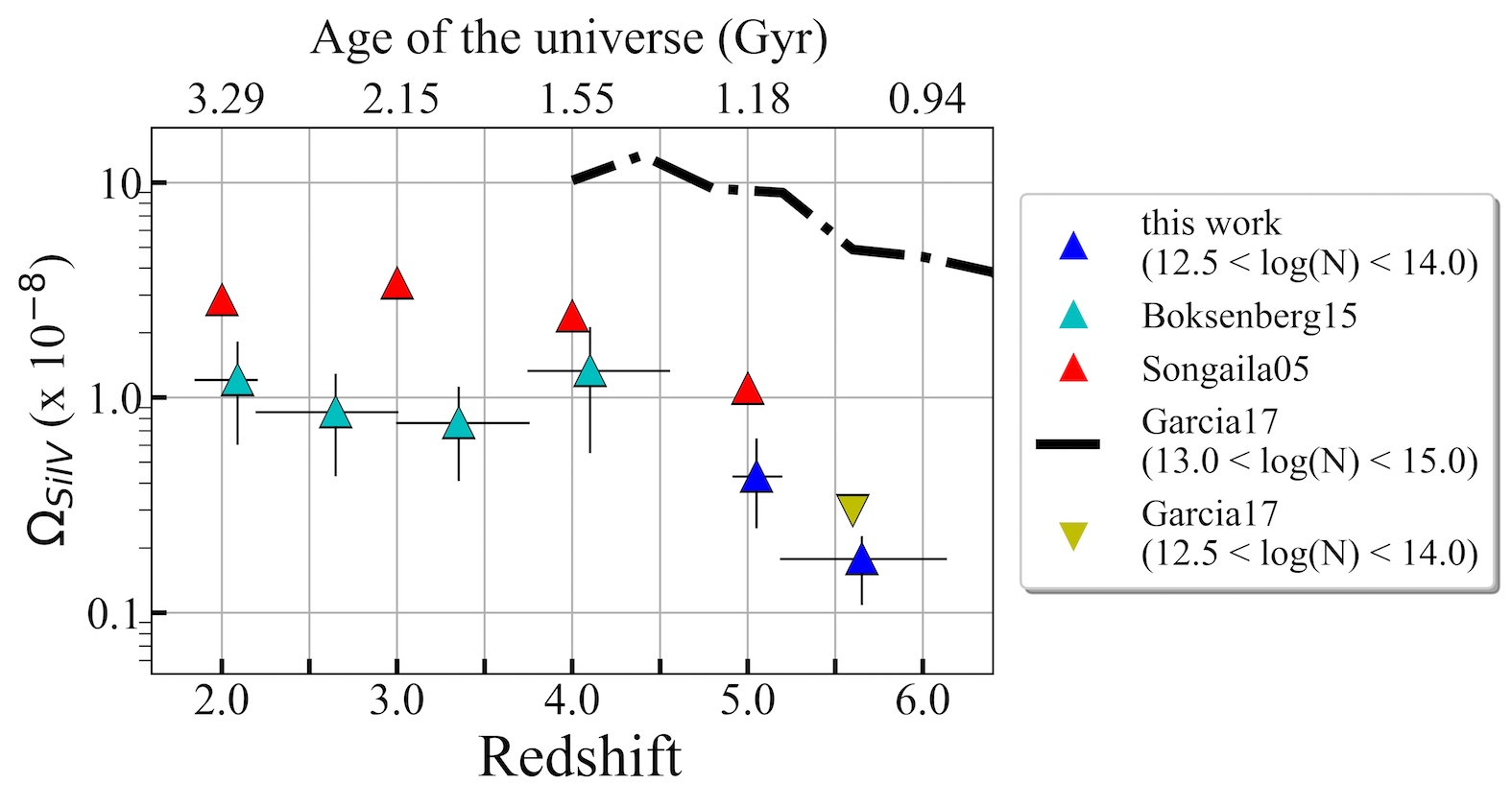}	
	 	 \caption{The evolution of the comoving mass densities of \ion{Si}{IV} as measured in this work along with with other observational values from literature and the simulation of G17. The marker style and colours are described in the legend. The B15 and S05 values have been adjusted to the {\it Planck} cosmology used in this work.}
		 
 \label{fig:omegasi4values}
\end{figure*}

We present, for the first time, the comoving mass density of \ion{Si}{IV} ($\Omega_{\ion{Si}{IV}}$) beyond redshift 5. The values of B15 and \citealt{SONGAILA2005} (S05) are adjusted to the {\it Planck} cosmology used in this work and can be seen, along with our values, in Figure \ref{fig:omegasi4values}. Given that we only have 2 \ion{Si}{IV} systems below redshift 5.19, we simply consider the lower and upper log(N$_{\textrm{sys}}$/cm$^{2}$) bounds of each component and compute the error boundary. We measure $\Omega_{\ion{Si}{IV}}$ = 4.3$^{+2.1}_{-2.1}$ $\times$10$^{-9}$ at a median redshift <z> = 5.05 and find that it drops to $\Omega_{\ion{Si}{IV}}$ = 1.4$^{+0.6}_{-0.4}$ $\times$10$^{-9}$ by the median redshift <z> = 5.66.

Interestingly, the G17 simulation, which accurately reproduces the evolution of $\Omega_{\ion{C}{IV}}$, also reproduces the observed relative evolution of $\Omega_{\ion{Si}{IV}}$. However, it overproduces \ion{Si}{IV} by an order of magnitude when considering the column density range 13.00 < log(N/cm$^{2}$) < 15.00 (dot-dash line Figure \ref{fig:omegasi4values}). If the comparison is restricted to the column density range of the systems discovered in this study (12.50 < log(N/cm$^{2}$) < 14.00) then they measure $\Omega_{\ion{Si}{IV}}$ = 3.04$\times$10$^{-9}$ which is within 3$\sigma$ of our measured values (see Table \ref{tab:incidenceomegatable}). We will discuss this in further detail in Sec. \ref{sec:si4discussion}.

\subsection{Column density distribution functions}
\label{sec:cddf} 

 We compute the respective CDDF for each ion in the following manner, \begin{equation}
 	f(\textrm{N}) = \frac{\textrm{n}}{\Delta log(\textrm{N}_{sys})} \times \frac{1}{dX} \\
	\label{eq:fn}
\end{equation}
\noindent where $N_{sys}$ is the total column density of a system, $n$ is the number of completeness corrected systems in the column density bin considered ($\Delta log(\textrm{N}_{sys})$) and $dX$ is the comoving redshift path (eq. \ref{eq:deltaxdistance}). The completeness functions associated with each redshift bin for \ion{C}{IV} and \ion{Si}{IV} can be seen in Fig. \ref{fig:cbarbyzbin}. We then perform a powerlaw fit to the column density distribution 

\begin{equation}
 	f(\textrm{N}) = B \times \Bigg( \frac{N_{sys}}{N_{0}} \Bigg)^{-\alpha} \\
	\label{eq:fnalpha}
\end{equation}
\noindent where $N_{0}$ = 10$^{13.64}$. We adopt this value for $N_{0}$ in order to compare to the work of D13. 

\begin{figure}
	\includegraphics[width = 8.4cm]{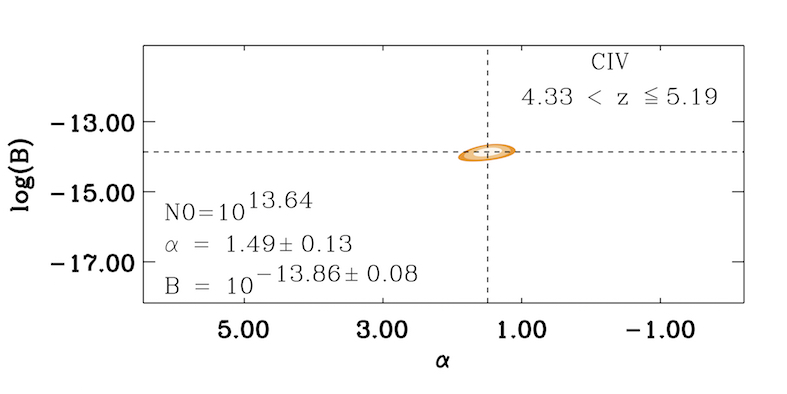}
	\caption{Example CDDF best fit 3$\sigma$ contour plots for all fit parameters (see eq. \ref{eq:fnalpha}) presented in Table \ref{tab:cddftable}. All CDDF best fits are presented in the online appendix.} 
	
	\label{fig:cddfplots}
\end{figure}

\begin{figure}
	\includegraphics[width=8.4cm]{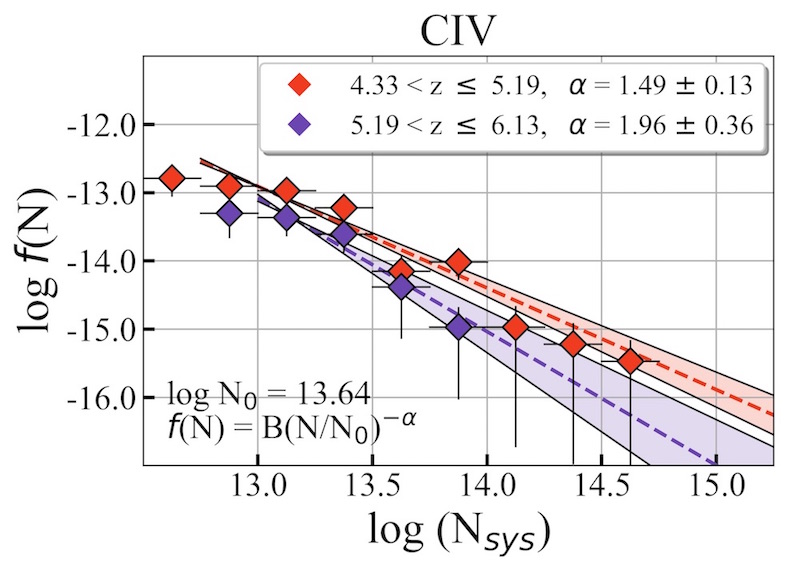}
	\caption{\ion{C}{IV} CDDF in two redshift bins. We use Poisson vertical error bars and the horizontal bars denote the column density range of each bin. The marker style and colours are described in the legend. All best fit values and 1$\sigma$ errors are presented in Table \ref{tab:cddftable}. The best fits are plotted with a dashed line and the 1$\sigma$ error bounds are shaded with colours consistent with those of the markers.} 
	
 \label{fig:c4cddf}
 \end{figure}

\begin{figure}
	\includegraphics[width=8.4cm]{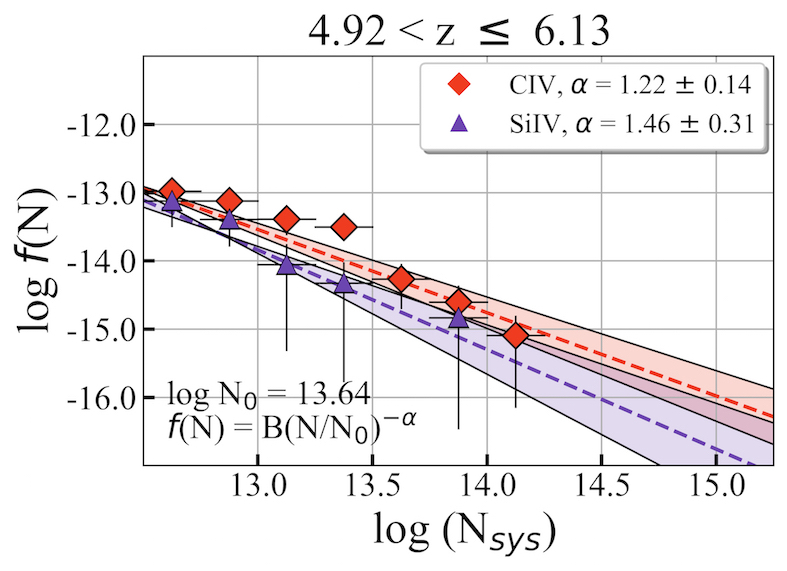}
	\caption{\ion{C}{IV} and \ion{Si}{IV} CDDF for systems in the same redshift bin. The {key to the} plotted values, errors and best fits {match} the description in Figure \ref{fig:c4cddf}.} 
	
	 \label{fig:si4c4cddf}
\end{figure}

\begin{figure}
	\includegraphics[width=8.4cm]{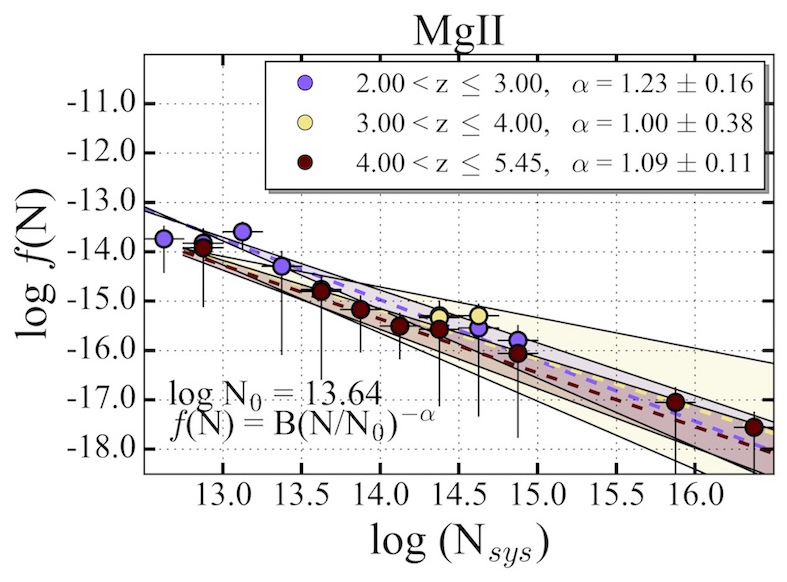}
	\caption{\ion{Mg}{II} CDDF in increasing redshift order. The {key to the} plotted values, errors and best fits {match} the description in Figure \ref{fig:c4cddf}.} 
	
 \label{fig:mg2cddf}
\end{figure}

\begin{figure}
	\includegraphics[width=8.4cm]{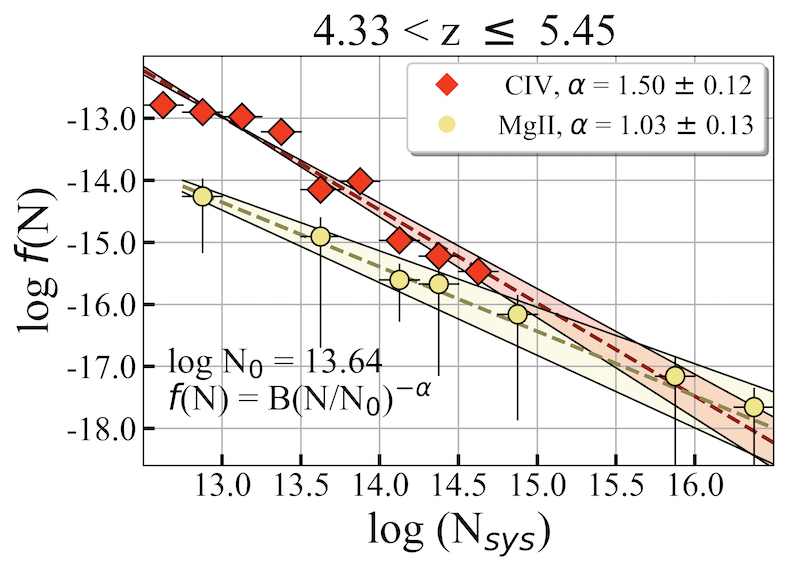}
	\caption{\ion{C}{IV} and \ion{Mg}{II} CDDF for systems in the same redshift bin. The {key to the} plotted values, errors and best fits {match} the description in Figure \ref{fig:c4cddf}.} 
	
	 \label{fig:mg2c4cddf}
\end{figure}

We follow the prescription put forth in B17 and take a maximum-likelihood expectation (MLE) approach. We simultaneously fit for $\alpha$ and $B$ and the likelihood function is defined as

\begin{equation}
 	L(\alpha, B) = P(n $ $| $ $\alpha, $ $B) \times \Pi P(N_{sys_{i}}$ $ | $ $\alpha) \\
	\label{eq:likelihood}
\end{equation}
\noindent where $P(n $ $| $ $\alpha, $ $B$) is the Poisson probability of observing $n$ systems given a single instance of $\alpha$ and $B$. We normalise $\Pi P(N_{sys_{i}} $ $| $ $\alpha)$ so that the expected number of systems is $n$. The redshift boundaries and transitions considered and best fit values with 1$\sigma$ errors are presented in Table \ref{tab:cddftable}. We also investigate if the choice of $N_{0}$ impacts our results as B17 used log(N$_{0}$) = 13.50 and find that the new values are well within the range of values presented in Table \ref{tab:cddftable}. {The contour plots of each fit are presented in Section H of the online appendix. An example can be seen in Figure \ref{fig:cddfplots}}. {The binned values in Figures \ref{fig:c4cddf}, \ref{fig:si4c4cddf}, \ref{fig:mg2cddf}, \ref{fig:mg2c4cddf}, \ref{fig:si4cddf} and \ref{fig:c4cddfonly} are presented for a by eye comparison to the best fits obtained through the MLE method discussed above.} 

\begin{table}
 \centering
 \caption{\ion{C}{IV}, \ion{Si}{IV} and \ion{Mg}{II} best MLE parameters for CDDF functional form (see eq. \ref{eq:fnalpha}). The redshift bins ($\Delta z$), column density range ($\Delta$log(N$_{\textrm{sys}}$/cm$^{2}$)), $\alpha$ and $B$ best fit values and 1$\sigma$ errors are presented below and are plotted in figures \ref{fig:c4cddf}, \ref{fig:si4c4cddf}, \ref{fig:mg2cddf} and \ref{fig:mg2c4cddf}. For consistency with D13, we use log(N$_{0}$) = 13.64. }
	\label{tab:cddftable}

 \begin{tabular}{|| c | c | c | c | c ||}
 \hline
Ion & $\Delta$ z & $\Delta$ $\log$(\textrm{N$_{\textrm{sys}}$/cm$^{2}$}) & $\alpha$ & $\log$(B) \\
\hline

 \ion{C}{IV} & 4.33-5.19 & [12.75, 14.75] & 1.49 $\pm$ 0.13 &-13.86 $\pm$ 0.08 \\
 	         & 5.19-6.13 & [13.00, 14.00] & 1.96 $\pm$ 0.36 &-14.33 $\pm$ 0.18 \\

 \ion{Mg}{II} & 2.00-3.00 & [12.50, 15.00] & 1.23 $\pm$ 0.16 &-14.54 $\pm$ 0.14 \\
 	          & 3.00-4.00 & [13.50, 15.00] & 1.00 $\pm$ 0.38 &-14.83 $\pm$ 0.33 \\
 	          & 4.00-5.45 & [12.75, 16.50] & 1.09 $\pm$ 0.11 &-14.98 $\pm$ 0.17 \\

 \ion{C}{IV}  & 4.33-5.45 & [12.50, 14.75] & 1.50 $\pm$ 0.12 &-13.94 $\pm$ 0.07 \\
 \ion{Mg}{II} &           & [12.75, 15.00] & 1.03 $\pm$ 0.13 &-15.03 $\pm$ 0.21 \\

 \ion{C}{IV}  & 4.92-6.13 & [12.50, 14.25] & 1.22 $\pm$ 0.14 &-14.29 $\pm$ 0.11 \\
 \ion{Si}{IV} &           & [12.50, 14.00] & 1.46 $\pm$ 0.31 &-14.77 $\pm$ 0.25 \\

 \end{tabular}

\end{table}


First, we fit the \ion{C}{IV} CDDF and the best fits and binned values can be seen in Figure \ref{fig:c4cddf}. We investigate the impact of the lowest column density considered as the \ion{C}{IV} CDDF appears to flatten below log(N$_{\textrm{sys}}$/cm$^{2}$) = 13.25. We exclude the lowest column density bin and do not find a significant impact on the final fit parameters. For example, when we {use} the full range of log(N$_{\textrm{sys}}$/cm$^{2}$) in the redshift range 4.33 < $z$ $\le$ 5.19 we find the best fit values $\alpha$ = 1.49 $\pm$ 0.13 and log(B) = -13.86 $\pm$ 0.08. If we limit the range to those \ion{C}{IV} systems with log(N$_{\textrm{sys}}$/cm$^{2}$) $\ge$ 12.75, we find the best fit values $\alpha$ = 1.69 $\pm$ 0.13 and log(B) = -13.79 $\pm$ 0.08. We find that the best fit $\alpha$ parameter increases to 1.96 $\pm$ 0.36 as we move beyond redshift 5.19. 

{Our values are within 1$\sigma$ of the D13 values } who measures $\alpha$ = 1.62 $\pm$ 0.2 (4.36 < $z$ < 5.3) and $\alpha$ = 1.44 $\pm$ 0.3 (5.3 < $z$ < 6.20). {A point of difference is that our column density fitting range does not extend beyond log(N$_{\textrm{sys}}$/cm$^{2}$) =14.00 in our highest redshift bin (5.19 < $z$ $\le$ 6.13) as in the D13 study.}


Next, we compare the \ion{C}{IV} and \ion{Si}{IV} CDDFs in the same redshift {range} across the redshift range for which they can be both observed: 4.92 < $z$ $\le$ 6.13. The results can be seen in Figure \ref{fig:si4c4cddf}. {For the \ion{Si}{IV} CDDF we compute the best fit values $\alpha$ = 1.46 $\pm$ 0.31 and log(B) = -14.77 $\pm$ 0.25 (see Table \ref{tab:cddftable})}. When considering the redshift boundary 5.19 < $z$ $\le$ 6.13 we find that the \ion{C}{IV} $\alpha$ best fit parameter does not {increase}, but rather {decreases} to $\alpha$ = 1.22 $\pm$ 0.14. This change is driven by the lack of systems with log(N$_{\textrm{sys}}$/cm$^{2}$) > 14.25. We see this fluctuation as reflecting the uncertainty associated with our small sample size but we do again highlight that, in all redshift bins, our best fit parameters our within 1$\sigma$ of the D13 values.


We also present, for the first time, the \ion{Mg}{II} CDDFs and best fits in three redshift bins in the full redshift range 2.0 < $z$ $\le$ 5.45. The results {are based on the 5$\sigma$ recovery selected systems reported in C17 and} can be seen in Table \ref{tab:cddftable} and Figure \ref{fig:mg2cddf}. We find that the best fit parameter $\alpha$ is consistent with $\alpha$ = 1.23 $\pm$ 0.16 in the redshift range 2.0 < $z$ < 3.0 and $\alpha$ = 1.09 $\pm$ 0.11 in the redshift range 4.0 < $z$ $\le$ 5.45. Finally, we compute the \ion{C}{IV} and \ion{Mg}{II} CDDFs across the redshift range for which they can be both observed: 4.33 < $z$ $\le$ 5.45. The binned values and best fits can be seen in Figure \ref{fig:mg2c4cddf}.


\section{Discussion}
\label{sec:discussion}

We measure, for the first time, the incidence rates and comoving mass density of \ion{Si}{IV} ($\Omega_{\ion{Si}{IV}}$) beyond redshift 5.5 and provide additional measurements on the incidence rates and comoving mass density of \ion{C}{IV} ($\Omega_{\ion{C}{IV}}$) beyond redshift 4.33. All values computed in this study are presented in Table \ref{tab:incidenceomegatable} and can be seen in Figures \ref{fig:dndzdX}, \ref{fig:omegac4values} and \ref{fig:omegasi4values}.

The statistics associated with the \ion{C}{IV} and \ion{Si}{IV} doublets are adjusted for the human impact (see Section \ref{sec:userinteraction}) on the output of an automated detection algorithm (see Section \ref{sec:visualcheck}). We adjust for the likely contamination by false positive detections and varying completeness across redshift bins (see Section \ref{sec:compfps}). We use the completeness adjusted values to compute (Equation \ref{eq:fn}) and fit (Equation \ref{eq:fnalpha}) the column density distribution functions (CDDFs) of the \ion{Si}{IV} and \ion{C}{IV} systems identified in this work as well as the that of the \ion{Mg}{II} systems provided in \cite{CODOREANU2017}. The best fit MLE parameters of the fitting functions are provided in Table \ref{tab:cddftable} and can be seen in Figures \ref{fig:c4cddf}, \ref{fig:si4c4cddf}, \ref{fig:mg2cddf} and \ref{fig:mg2c4cddf}.

Our study is the first to provide a comprehensive survey {which compares} both \textit{low} and \textit{high ionisation} systems beyond redshift $z$ = 5. Below we compare to the the fiducial model of \cite{GARCIA2017} (G17); we also discuss the physical connection between \textit{low} and \textit{high ionisation} absorbers.

\subsection{SiIV systems}
\label{sec:si4discussion}

{First, we focus on the} \ion{Si}{IV} {systems identified in this work}. As we have seen in Section \ref{sec:omegasi4}, and as expected, the column density range of the CDDF impacts the computed $\Omega_{\ion{Si}{IV}}$. The {results from the} simulation of G17 are in remarkable agreement when {restricted to} the column density range of \ion{Si}{IV} systems discovered in this study (12.50 $\le$ log(N$_{\textrm{sys}}$/cm$^{2}$) $\le$ 14.0; see Table \ref{tab:cddftable}) but differs by approximately an order of magnitude when we {compare to} the column density range 13.00 $\le$ log(N$_{\textrm{sys}}$/cm$^{2}$) $\le$ 15.00 (dot-dash line in Figure \ref{fig:omegasi4values}).

\begin{figure}
	\includegraphics[width=8.4cm]{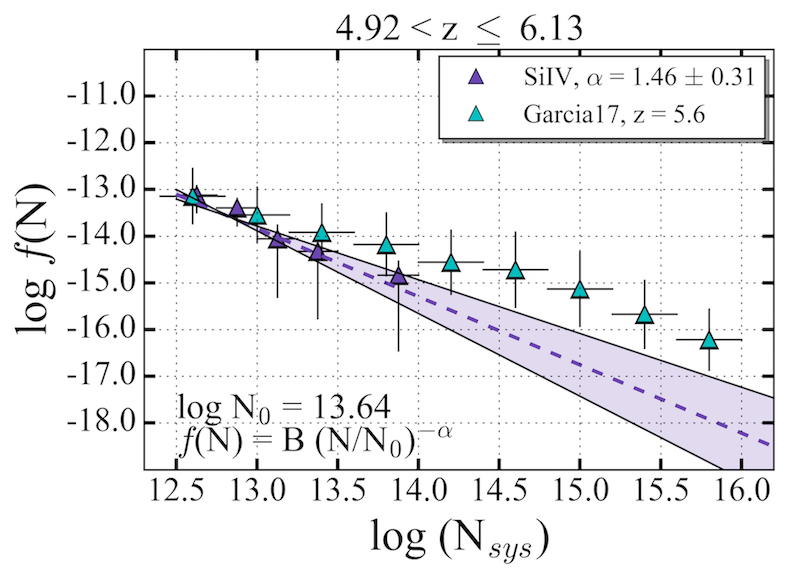}
	\caption{\ion{Si}{IV} CDDF for systems in the redshift bin 4.92 $<$ z $\le$ 6.13. The {key to the} plotted values, errors and best fits {match} the description in Figure \ref{fig:c4cddf}.} 

	 \label{fig:si4cddf}
\end{figure}

In order to investigate this behaviour, we plot our CDDF values and best fits with the binned CDDF values from G17 (see Figure \ref{fig:si4cddf}). As can be observed, the two distributions are in good agreement only up to log(N$_{\textrm{sys}}$/cm$^{2}$)$\sim$14.00. For values of log(N$_{\textrm{sys}}$/cm$^{2}$) > 14.00, the \ion{Si}{IV} CDDF of G17 disagrees with the extrapolation of the functional fit to the observed distribution. {In order to investigate whether the \ion{Si}{IV} CDDF of G17 could be representative of the true \ion{Si}{IV} population, we ask how many \ion{Si}{IV} systems with log(N$_{\textrm{sys}}$/cm$^{2}$) >14.00 we should have observed over the absorption path of our survey ($\Delta$X$_{\textrm{\ion{Si}{IV}}}$ = 16.40; see Table \ref{tab:incidenceomegatable})}.

We consider the CDDF of G17 {in the column density range 14.4 $\le$ log(N$_{\textrm{sys}}$/cm$^{2}$) $\le$ 14.8}. {We} find that we should have detected $\sim$12 \ion{Si}{IV} absorbers{within the column density range considered. We observe none. Next,} we consider the case that the CDDF fit of this work extends past log(N$_{\textrm{sys}})\sim$14.00 and use the best fit values in Table \ref{tab:cddftable} for the redshift range 4.92 $\le$ z $\le$ 6.13. When considering the {same} column density range we calculate an expectation of $\sim$0.4 absorbers. This suggests that future surveys which will increase $\Delta$X$_{\textrm{\ion{Si}{IV}}}$ to beyond $\sim$40 should observe a \ion{Si}{IV} system with 14.4 $\le$ log(N$_{\textrm{sys}}$/cm$^{2}$) $\le$ 14.8. Future surveys {with enough path length to at least detect one such system} will be able to better address this issue.

{While we do not detect any systems with column densities between 14.00 < log(N$_{\textrm{sys}}$/cm$^{2}$) < 15.00 we can investigate their impact on the associated $\Omega$ value. We integrate the first moment of the \ion{Si}{IV} CDDF and compute the resulting $\Omega_{\ion{Si}{IV}}$ value (see Equation \ref{eq:omegaintegral}) in the redshift range 4.92 $<$ z $\le$ 6.13 and column density range 12.50 < log(N$_{\textrm{sys}}$/cm$^{2}$) < 15.00. We compute an expectation of $\Omega_{\ion{Si}{IV}}$ = 5.6$^{+11.7}_{-3.5}$ $\times$10$^{-9}$. Next, we combine the $\Omega_{\ion{Si}{IV}}$ values measured in this study from Table \ref{tab:incidenceomegatable} and find that across the same redshift bin we measure $\Omega_{\ion{Si}{IV}}$ = 5.7$^{+2.7}_{-2.5}$ $\times$10$^{-9}$. This suggests that the contribution to $\Omega_{\ion{Si}{IV}}$ from future detections of \ion{Si}{IV} with 14.00 < log(N$_{\textrm{sys}}$/cm$^{2}$) < 15.00 will be minimised by the large absorption path necessary to discover them. }

{Given then that our measured $\Omega_{\ion{Si}{IV}}$ value is not biased by the lack of \ion{Si}{IV} systems with log(N$_{\textrm{sys}}$/cm$^{2}$) > 14.00, we next look at the relative evolution of \ion{Si}{IV} from below to above z$\sim$5. Our motivation for this is to compare to the relative evolution of $\Omega_{\ion{C}{IV}}$ which drops by a factor of 2 to 4 across the same redshift range (D13 and references therein). We adjust the highest redshift values from B15 {to} the cosmology used in this paper and find $\Omega_{\ion{Si}{IV}}$ = 1.4 $\pm$ 0.8$\times$10$^{-8}$ at <z> = 4.10. When we compare to the $\Omega_{\ion{Si}{IV}}$ measured in this study (see Table \ref{tab:incidenceomegatable}), we find that $\Omega_{\ion{Si}{IV}}$ drops by a factor of 3 $\pm$ 2 from <z> = 4.10 to <z> = 5.05. By <z> = 5.66 that factor increases to 10 $\pm$ 7. }

{For comparison we calculate the fractional evolution of $\Omega_{\ion{C}{IV}}$ using systems identified in this work (see Table \ref{tab:incidenceomegatable}) and find that $\Omega_{\ion{C}{IV}}$ decreases by a factor of 5$\pm$2 from <z> = 4.77 to <z> = 5.66. While this seems to suggest that $\Omega_{\ion{Si}{IV}}$ and $\Omega_{\ion{C}{IV}}$ have similar evolutions from below to {beyond} redshift 5 we again highlight that our work is the only study, so far, to provide information on $\Omega_{\ion{Si}{IV}}$ beyond redshift 5.5. Future studies will be able to provide further insight on this evolution.}

\begin{figure}
	\includegraphics[width=8.4cm]{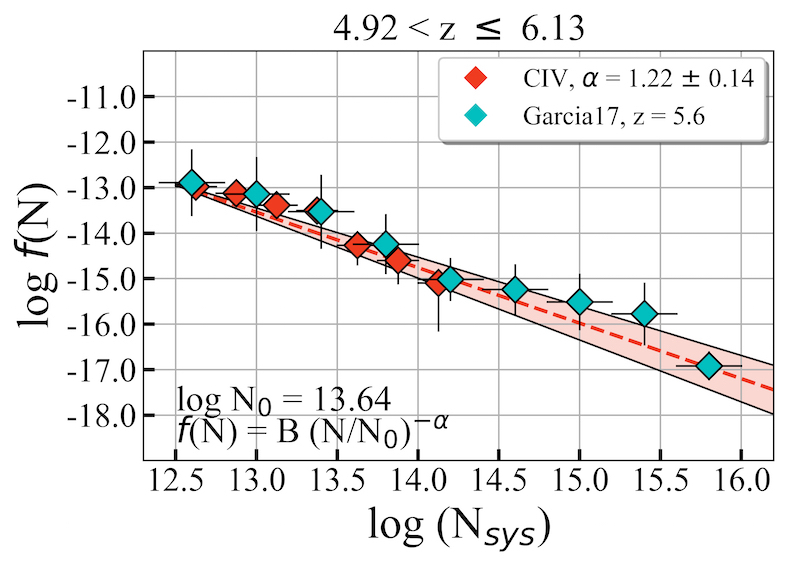}
	\caption{\ion{C}{IV} CDDF for systems in the redshift bin 4.92 $<$ z $\le$ 6.13. The {key to the} plotted values, errors and best fits {match} the description in Figure \ref{fig:c4cddf}.} 

	 \label{fig:c4cddfonly}
\end{figure}

Interestingly, five of the seven \ion{Si}{IV} systems also have associated \ion{C}{IV}. All of these \ion{Si}{IV} systems have log(N$_{\textrm{sys}}$/cm$^{2}$) $\ge$ 12.75. Two of these systems, 8 and 9 in $ULAS$ $J1319+0959$, also have associated \textit{low ionisation} systems, \ion{Mg}{II}+\ion{Al}{II}($\lambda$1670) and \ion{C}{II}($\lambda$1334)+\ion{Al}{II}($\lambda$1670) respectively. As \cite{FINLATOR2016} have pointed out, \ion{Si}{IV} ($\sim$3.3 Ry) is an intermediary transition between \ion{C}{II} (1.8 Ry) and \ion{C}{IV} ($\sim$4.7 Ry). Thus, constraining both the relative evolution of \ion{Si}{IV}, \ion{C}{IV} and \ion{C}{II} as well as the number density and nature/environment of \ion{Si}{IV} and \ion{C}{IV} systems with log(N$_{\textrm{sys}}$/cm$^{2}$) > 14 is necessary in order to discriminate between different UVB or hydrogen self-shielding prescriptions. Given that we do not identify any \ion{Si}{IV} systems with log(N$_{\textrm{sys}}$/cm$^{2}$) > 14 and that all \ion{Si}{IV} systems with log(N$_{\textrm{sys}}$/cm$^{2}$) > 12.75 identified in this study have associated \ion{C}{IV} we next investigate the population of \ion{C}{IV} absorbers {identified in this work.}

\subsection{CIV systems}
\label{sec:c4discussion}

As we have seen in {section} \ref{sec:cddf}, {and as expected ,} the CDDF slope is sensitive to both the sample size and the column density range of the fit. For example, the best fit parameter $\alpha$ {increases from 1.22 $\pm$ 0.14 to 1.96 $\pm$ 0.36 when the column density range changes from $\Delta$log(N$_{\textrm{sys}}$/cm$^{2}$) = [12.50, 14.25] to $\Delta$log(N$_{\textrm{sys}}$/cm$^{2}$) = [13.00, 14.00] and sample size decreases from 13 to 6 systems (see figures \ref{fig:c4cddf} and \ref{fig:si4c4cddf} respectively)}.

As in the previous section, we compare the \ion{C}{IV} CDDF and best fits computed in this work with the \ion{C}{IV} G17 CDDF. The results can be seen in Figure \ref{fig:c4cddfonly}. We find that both CDDFs are in excellent agreement in the column density range of the systems discovered in this work $\Delta$log(N$_{\textrm{sys}}$/cm$^{2}$) = [12.50, 14.25]. Unlike the \ion{Si}{IV} CDDFs discussed in the previous section, the \ion{C}{IV} CDDF of G17 is in excellent agreement with an extrapolation of the functional fit to the \ion{C}{IV} CDDF computed in this work, log(N$_{\textrm{sys}}$/cm$^{2}$) > 14.25.

{As we have discussed in the previous section, the comoving mass density is sensitive to the column density {range} considered. We quantify this influence by computing the fractional contribution, $f (\Omega_{\ion{C}{IV}})$, from each column density bin, {$\Delta$ log(N$_{{\rm CIV}}$)}, considered in section \ref{sec:cddf} to the total computed $\Omega_{\ion{C}{IV}}$}:

\begin{equation}
 	f (\Omega_{CIV}) = \frac{\Omega_{\Delta \log(N_{CIV})}}{\Omega_{\ion{C}{IV}}} \\
	\label{eq:fomega}
\end{equation}

\begin{figure}
	\includegraphics[width=8.4cm]{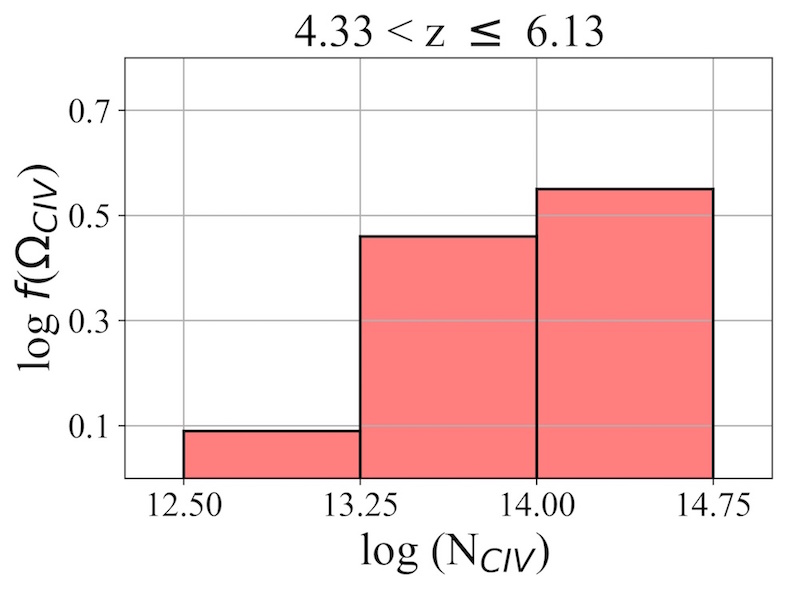}
	\caption{The {fractional contribution to} $\Omega_{\ion{C}{IV}}$ as measured in this study in $\Delta$log(N$_{\ion{C}{IV}}$) bins as defined in Equation \ref{eq:fomega}.}
	 \label{fig:c4fraction}
\end{figure}

\noindent We find that \ion{C}{IV} systems with log(N$_{\textrm{sys}}$/cm$^{2}$) $\ge$ 14.00 contribute more than 50$\%$ to the {comoving mass density} of \ion{C}{IV} {computed} in this work, $\Omega_{\ion{C}{IV}}$. The resulting histogram can be seen in Figure \ref{fig:c4fraction}.

In order to further investigate \ion{C}{IV} systems with log(N$_{\textrm{sys}}$/cm$^{2}$) $\ge$ 14.00, we subdivide the \ion{C}{IV} absorber population used in this study in 4 categories based on whether or not associated absorbers are also identified {at the redshift of the \ion{C}{IV} doublet}. We find: 

$\bullet$ 25 \ion{C}{IV} systems \textit{with no associated absorbers}

$\bullet$ 6 \ion{C}{IV} systems \textit{with associated low ionisation absorbers}: system 3 in $ULAS$ $J0148+0600$, system 5 in $SDSS$ $J0927+2001$, systems 2, 8 and 9 in $SDSS$ $J1306+0356$ and system 3 in $ULAS$ $J1319+0959$

$\bullet$ 2 \ion{C}{IV} systems \textit{with associated low $\&$ high ionisation absorbers}: systems 8 and 9 in $ULAS$ $J1319+0959$

$\bullet$ 3 \ion{C}{IV} systems \textit{with associated high ionisation absorbers}: system 7 in $ULAS$ $J0148+0600$, system 8 in $SDSS$ $J0927+2001$ and system 6 in $ULAS$ $J1319+0959$

\begin{figure}
	\includegraphics[width=8.4cm]{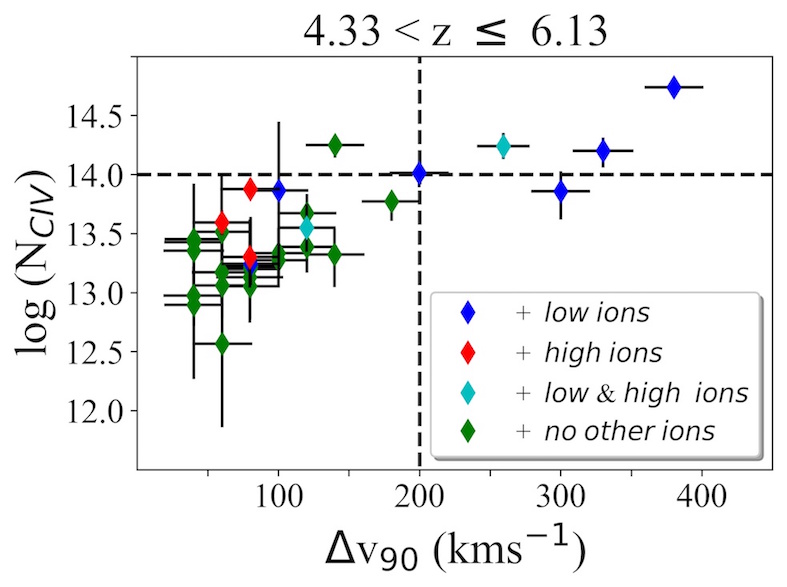}
	\caption{The velocity width ($\Delta$v$_{90}$) vs the column density, log(N$_{\textrm{\ion{C}{IV}}}$) of the \ion{C}{IV} systems used in this work. The marker style and colours are described in the legend. The labels are described in detail in Section \ref{sec:c4discussion}.}
	 \label{fig:c4v90}
\end{figure}

{Following this}, we compute the velocity width of each absorber ($\Delta$v$_{\textrm{90}}$; \citealt{PROCHASKA2008}) and plot the values vs. the column density of each system in Figure \ref{fig:c4v90}. We find that four out of the six \ion{C}{IV} systems with log(N$_{\textrm{sys}}$/cm$^{2}$) $\ge$14.00 have associated \textit{low ionisation} absorbers. Furthermore, we find that {all} \ion{C}{IV} systems with $\Delta$v$_{\textrm{90}}$ $\ge$ 200 kms$^{-1}$ have associated \textit{low ionisation} absorbers {(five of the eight absorbers)}. In order to highlight these boundaries, we have plotted dashed lines in Figure \ref{fig:c4v90}.

These findings suggest that \ion{C}{IV} absorbers {with} associated \textit{low ionisation} absorbers contribute significantly to the number density and the comoving mass density of \ion{C}{IV}. Furthermore, these same \ion{C}{IV} systems are dominated by a sub-population of absorbers with broad velocity profiles ($\Delta$v$_{\textrm{90}}$ $\ge$ 200 kms$^{-1}$). Next, we focus our discussion on the two absorbers with the broadest velocity profiles, systems 8 and 9 in $SDSS$ $J1306+0356$.

\subsection{The physical connection between low and high ionisation absorbers}
\label{sec:sys8and9discussion}
\begin{figure*}
	\includegraphics[width=14cm]{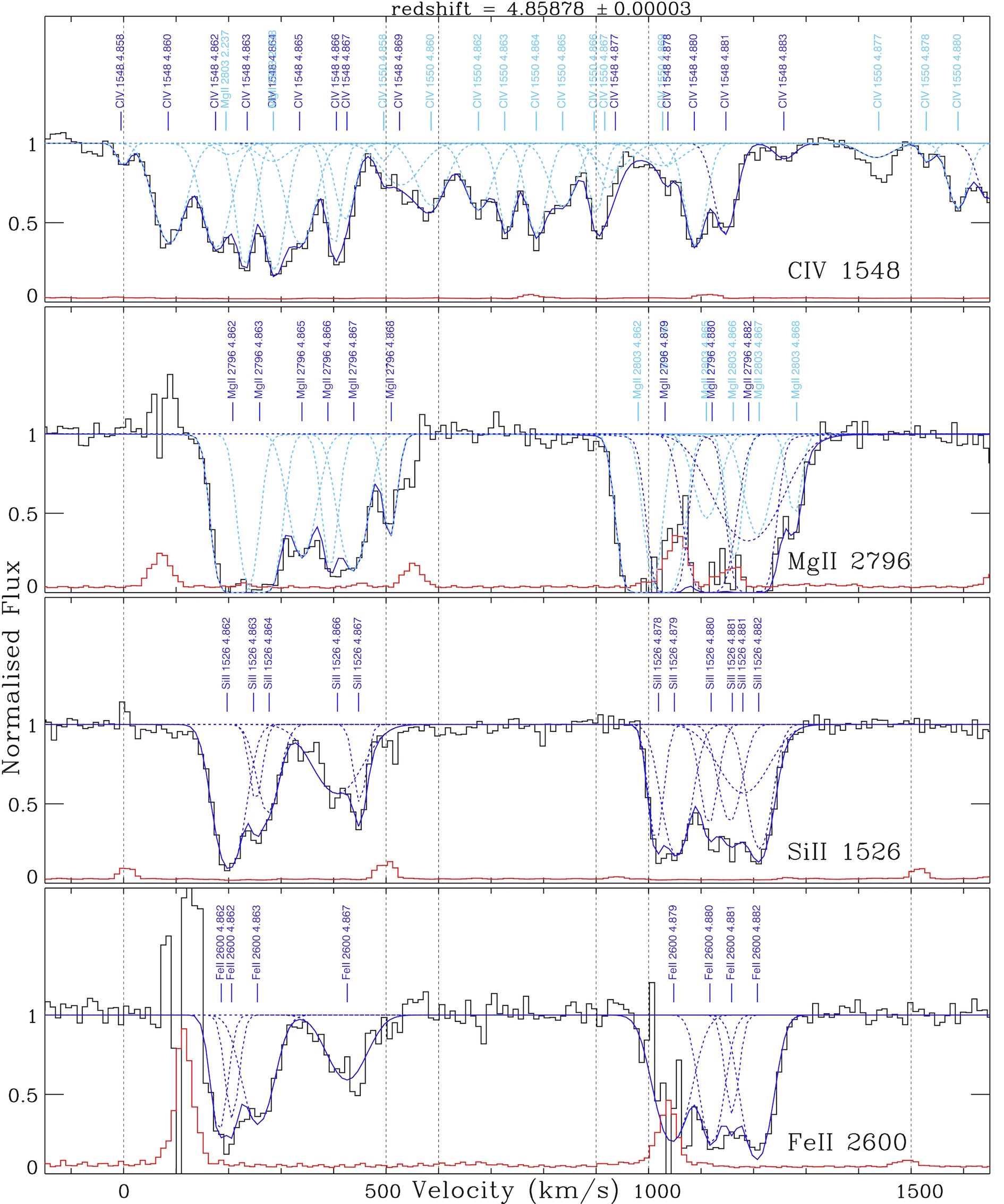}
	\caption{Selected transitions from systems 8 and 9 identified in the \textit{SDSS J1306+0356} sightline. Each transition is identified in the bottom middle of each panel. In each panel, the vertical axis is the continuum normalised flux. The horizontal axis is the velocity separation ( kms$^{-1}$) from the lowest redshift component of a system. The normalised spectrum is plotted in black and the associated error is in red. The solid blue line represents the full fit to the spectra and includes other ions besides the transition identified in the bottom right of each panel (i.e. the \ion{C}{IV}$\lambda$1550 transition). Individual components are plotted with dashed lines and are identified by a vertical label. The identified transition components are in solid blue and other transitions are in light blue. The black vertical dashed lines highlight the 0, +500, +600, +900, +1000 and the +1500 kms$^{-1}$ velocity locations for reasons discussed in Sec. \ref{sec:sys8and9discussion}.}	
	
 \label{fig:S1306_z_4d85878}
\end{figure*}

As we have previously discussed in section \ref{sec:s1306}, we have plotted selected transitions\footnote{\ion{C}{IV} ($\lambda$1548), \ion{Mg}{II}($\lambda$2796), \ion{Si}{II}($\lambda$1526) and \ion{Fe}{II}($\lambda$2600)} of systems 8 and 9 together in Figure \ref{fig:S1306_z_4d85878} in order to highlight the proximity and similar velocity structure of the associated absorbers.

{Our first question is to {assess} if the two systems could be associated with a single galaxy. In order to investigate this possibility, we translate the velocity separation between the reddest \ion{C}{IV} component of system 8 and {bluest} \ion{Mg}{II} component of system 9 (component $a$ in Table A3 in C17) into a physical separation. This separation is highlighted by the black vertical dashed lines at +600 and +900 kms$^{-1}$ in Fig. \ref{fig:S1306_z_4d85878} which correspond to redshifts 4.87051 and 4.87638, respectively. The physical distance between the two redshifts is $\sim$550 kpc\footnote{$\sim$3.23 comoving Mpc}. Given that the systems themselves span more than 500 kms$^{-1}$ and are separated by $\sim$550 physical kpc we find that these two absorption systems are most likely tracing two separate structures rather a fortuitous double intersection of our sightline through a single system. }

Understanding the environment of such \ion{C}{IV} and \ion{Mg}{II} systems is a crucial step as they have a significant impact on the associated CDDFs and comoving mass densities. Previous studies which have investigated similar \ion{Mg}{II} absorbers\footnote{with W$_{2796}$ > 3 \AA$ $} at z < 1 have found that such ultra strong absorbers reside in group environments. The velocity width of the absorbers can be driven by either star formation (SFR > 5 M$_{\sun}$yr$^{-1}$) driven outflows \citep{NESTOR2011} or intra-group interactions \citep{GAUTHIER2013}. {Furthermore, such strong \ion{C}{IV} systems (log(N$_{sys}$/cm$^2$)>14.0) have also been connected to dense environments populated with young and blue galaxies in the redshift range 1.8$\le$ $z$ $\le$3.3 \citep{ADELBERGER2005}. Are then these systems tracing similar physical environments past redshift 5?}

Recently, \citet{CAI2017} also investigated the environment of these same \ion{C}{IV} absorbers using HST {narrow band} imaging to identify Ly$\alpha$ emitters (LAE). They {find} a {single} LAE candidate at an impact parameter of 205 kpc with an associated Ly$\alpha$ luminosity derived star formation rate, SFR$_{\textrm{Ly}_{\alpha}}$ = 2.5 M$_{\sun}$yr$^{\textrm{-1}}$. This would suggest that there are no star forming galaxies {with SFR > 5 M$_{\sun}$yr$^{-1}$} which can be associated with the absorber.

{Given that the study of \citet{CAI2017} does not identify multiple star forming galaxies we favour} the hypothesis that the velocity width of the absorbers is driven by tidal interactions {between previous outflow material and group members with luminosities below the detection limit of the \citet{CAI2017} study. We find that these \ion{Mg}{II} systems likely trace a disturbed environment, a possibility recently raised by \cite{ZOU2017} following their study of strong \ion{Mg}{II} systems in the redshift range 1.73<z<2.43.} However, since we currently do not have any information on the associated galaxies, it is difficult to create a complete picture.

Despite {this} lack of information, the physical association of \ion{C}{IV} systems with log(N$_{\textrm{sys}}$/cm$^{2}$) $\ge$14.00 with \textit{low ionisation} systems and their high velocity width ($\Delta$v$_{90}$ $\ge$ 200 kms$^{-1}$) suggests that these systems are tracing a multi-phase medium where hot and cold gas is mixing at the interface between the CGM and IGM. Thus, in order to accurately simulate the full \ion{C}{IV} population of absorbers this physical interaction must be accounted for.

\subsection{The comoving mass densities of low ionisation systems beyond redshift $\sim$2}
\label{sec:highlow}

\begin{figure}
	\includegraphics[width=8.6cm]{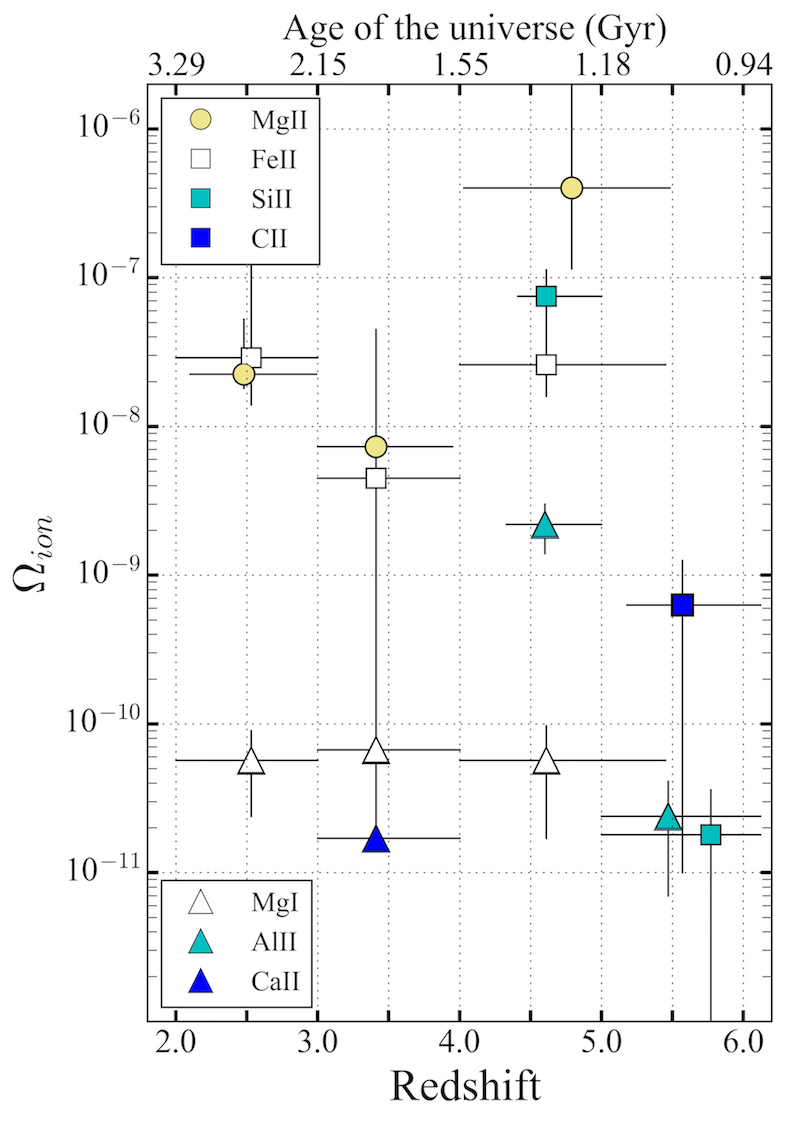}	
	 \caption{The evolution of the comoving mass densities of \ion{Mg}{I}, \ion{Ca}{II}, \ion{Si}{II}, \ion{Al}{II}, \ion{Fe}{II} and \ion{C}{II} as measured in this study. The redshift boundaries, median redshifts and computed $\Omega_{ion}$ values along with errors can be seen in Table \ref{tab:lowions}. The \ion{Mg}{II} values are taken from C17. }

 \label{fig:omegavalues}
\end{figure}
We have so far provided further measurements of $\Omega_{\ion{C}{IV}}$ (see Section \ref{sec:omegac4} and Figure \ref{fig:omegac4values}) and presented the first measurement of $\Omega_{\ion{Si}{IV}}$ (see Section \ref{sec:omegasi4} and Figure \ref{fig:omegasi4values}) beyond redshift 5. However, as we have discussed in Section \ref{sec:visualcheck}, we also search for other associated transitions at the redshift of the identified doublets. {We do not identify any \ion{O}{I} systems over the absorption path of our survey ($dX$ = 8.93).}

When considering the \ion{C}{IV} and \ion{Si}{IV} doublets identified in our work, we find 5 \ion{Si}{II}, 8 \ion{Al}{II}, 6 \ion{Fe}{II}, 1 \ion{C}{II} and 2 \ion{Mg}{I} associated transitions with z > 4.33. These transitions are associated with:

- systems 3 and 9 in sightline \textit{ULAS J0148+0600}

- system 5 in sightline \textit{SDSS J0927+2001} 

- systems 2, 8 and 9 in sightline \textit{SDSS J1306+0356}

- systems 8 and 9 in sightline \textit{ULAS J1319+0959}

\noindent We find a further 2 \ion{Al}{II}, 6 \ion{Fe}{II}, 5 \ion{Mg}{I} and 1 \ion{Ca}{II} associated with \ion{Mg}{II} doublets below redshift 4.33. We present these absorbers in {the online appendix}.

All of these transitions are associated with doublet identified absorbers which have a $user$ $success$ and $failure$ adjusted recovery rates greater than 50$\%$ (see Section \ref{sec:surveycompandfp}). As such, we consider these identifications to be robust. We do not adjust their statistics {for completeness} and compute their associated comoving mass densities using eq. \ref{eq:omegaion}. The errors are calculated using {Poisson statistics}. We present the comoving mass densities of \ion{Fe}{II}, \ion{Si}{II}, \ion{C}{II}, \ion{Mg}{I}, \ion{Al}{II} and \ion{Ca}{II} in Table \ref{tab:lowions} and Figure \ref{fig:omegavalues}. We compare our results with the recent work of \cite{WENLAN2017} (LF17) which provides the comoving mass densities of these ions along with others not identified in this work up to redshift $\sim$2.5 as traced by \ion{Mg}{II} absorbers. 
\begin{table}
 \centering
 \caption{Median redshifts (<z>), redshift paths ($\Delta$z), absorption distances (dX), discovered systems ($\bar{N}$) and the comoving mass densities ($\Omega$) for \ion{C}{II}, \ion{Si}{II}, \ion{Al}{II} and \ion{Fe}{II} systems used in this work.}
 \label{tab:lowions}
 \begin{tabular}{|| c | c | c | c | c | c ||}
 \hline

 ion & <z> & $\Delta$z & dX & $\bar{N}$ & $\Omega$\\

 \hline
 
\ion{Si}{II} & 4.61 & 4.41-5.00 & 7.70 & 4 & 7.5 $\pm$ 3.8$\times$10$^{-8 }$ \\
& 5.77 & 5.00-6.13 & 17.59 & 1 & 1.8 $\pm$ 1.8$\times$10$^{-11}$ \\

\ion{Al}{II} & 4.60 & 4.00-5.00 & 16.78 & 8 & 2.2 $\pm$ 0.8$\times$10$^{-9 }$ \\
 & 5.47 & 5.00-6.13 & 17.59 & 2 & 2.4 $\pm$ 1.7$\times$10$^{-11}$ \\
 
\ion{Fe}{II} & 2.53 & 2.00-3.00 & 13.12 & 4 & 2.9 $\pm$ 1.5$\times$10$^{-8 }$ \\
 & 3.41 & 3.00-4.00 & 15.09 & 1 & 4.5 $\pm$ 4.5$\times$10$^{-9 }$ \\
 & 4.61 & 4.00-5.45 & 24.84 & 7 & 2.6 $\pm$ 1.0$\times$10$^{-8 }$ \\

\ion{C}{II} & 5.57 & 5.18-6.13 & 11.32 & 1 & 6.3 $\pm$ 6.3$\times$10$^{-10}$ \\

\ion{Mg}{I} & 2.53 & 2.00-3.00 & 13.12 & 3 & 5.7 $\pm$ 3.3$\times$10$^{-11}$ \\
 & 3.41 & 3.00-4.00 & 15.09 & 2 & 6.7 $\pm$ 4.7$\times$10$^{-11}$ \\
 & 4.61 & 4.00-5.45 & 24.84 & 2 & 5.7 $\pm$ 4.0$\times$10$^{-11}$ \\

\ion{Ca}{II} & 3.41 & 3.00-4.00 & 15.09 & 1 & 1.7 $\pm$ 1.7$\times$10$^{-11}$ \\

 \end{tabular}
\end{table}

We find that our $\Omega_{\ion{Fe}{II}}$ and $\Omega_{\ion{Mg}{I}}$ values are in excellent agreement with those presented in LF17 at z$\sim$2.5 and exhibit a flat evolution (within 2$\sigma$) from redshift 2 to 5.45. This is unsurprising as the \ion{Fe}{II} and \ion{Mg}{I} transitions are mostly\footnote{except for system 2 in \textit{SDSS J1306+0356} which has W$_{2796}$ = 0.734 $\pm$ 0.062 \AA $ $ \label{f1}} associated with strong \ion{Mg}{II}$ $ absorbers whose $\Omega_{\ion{Mg}{II}}$ exhibits a similar behaviour as discussed in C17. We find a similar flat evolution when comparing the value of $\Omega_{\ion{Si}{II}}$ computed in this work at <z> = 4.61 (7.5 $\pm$ 3.8$\times$10$^{-8}$) with those of LF17 at z$\sim$2.4 ($\sim$3$\times$10$^{-8}$). Again, this is not surprising since these \ion{Si}{II} and \ion{Fe}{II} transitions are associated with the same systems.

Interestingly, we find that $\Omega_{\ion{Si}{II}}$ drops by $\sim$3 orders of magnitude from <z> = 4.61 (7.5 $\pm$ 3.8$\times$10$^{-8}$) to <z> = 5.77 (1.8 $\pm$ 1.8$\times$10$^{-11}$). We find only one \ion{Si}{II} system {beyond} redshift 5 (z = 5.77495 $\pm$ 0.00038; see Figure \ref{fig:U0148b}) and it is anchored by a \ion{Si}{IV} doublet. However, just as the \ion{Fe}{II} and \ion{Mg}{I} systems discussed above, all other \ion{Si}{II} identified in this work are mostly\textsuperscript{\ref{f1}} associated with $strong$ \ion{Mg}{II} systems. We see the evolution of $\Omega_{\ion{Si}{II}}$ from <z> = 4.61 to <z> = 5.77 as simply reflecting this association, or lack there of, and further observational studies will be able to confirm this.

Similar to the evolution of \ion{Si}{II} from redshift 2 to 5.47, we find that $\Omega_{\ion{C}{II}}$ also drops by several orders of magnitude from $\sim$0.5$\times$10$^{-7}$ at z$\sim$2.4 (LF17) to 6.3 $\pm$ 6.3$\times$10$^{-9}$ at <z> = 5.57 as measured in this work. However, the single \ion{C}{II} transition identified in this work is associated with a single \textit{weak} \ion{Mg}{II} absorber while LF17 integrate across the full equivalent width range. When we consider the evolution of \ion{Al}{II} we find that it drops by $\sim$2 orders of magnitude from redshift 4.60 to 5.47. Just as with with the \ion{Si}{II} and \ion{C}{II} ions, the \ion{Al}{II} absorbers beyond redshift 5 are associated only with \textit{weak} \ion{Mg}{II} absorbers while the \ion{Al}{II} absorbers below redshift 4 are mostly associated with $strong$ \ion{Mg}{II} absorbers. These absorbers dominate the computed $\Omega_{\ion{Al}{II}}$ at <z> = 4.60.

We find that $\Omega_{\ion{Mg}{I}}$ and $\Omega_{\ion{Fe}{II}}$ have a flat evolution from redshift 2 to 5.45 as measured in this work. When we compare to the findings of LF17, we find a similar evolution for $\Omega_{\ion{Si}{II}}$ and $\Omega_{\ion{Al}{II}}$ from redshift 2 to 5. However, from redshift 5 to redshift 6 $\Omega_{\ion{Si}{II}}$, $\Omega_{\ion{C}{II}}$ and $\Omega_{\ion{Al}{II}}$ drop by several orders of magnitude. This evolution results from the association of these absorbers with only \textit{weak} \ion{Mg}{II} absorbers while, from redshift 2 to 5, the \ion{Si}{II} and \ion{Al}{II} identified in this study are mostly associated with $strong$ \ion{Mg}{II} absorbers. {For this reason, we caution the reader in drawing significant conclusions from their evolution as the \ion{Si}{II}, \ion{Ca}{II} and \ion{Al}{II} populations identified in this work past redshift 5 are clearly limited by small number statistics.} We will further investigate the full population and statistics of \ion{Mg}{II} absorbers from redshift 2 to 7 in an upcoming paper.

\section{Conclusion}
\label{sec:conclusion}

We investigate four medium resolution and signal-to-noise spectra of z$\sim$6 QSOs for the presence of \ion{C}{IV}, \ion{Si}{IV} doublets and associated transitions. These same spectra were investigated for the presence of \ion{Mg}{II} doublets in \cite{CODOREANU2017}. 

We adjust the statistics of the \ion{C}{IV} and \ion{Si}{IV} systems for the impact of varying signal-to-noise and completeness across the redshift bins considered, the human impact on the identification methodology and false positive contamination. The details are described in Section \ref{sec:surveycompandfp}. The incidence rates, absorption paths and comoving mass density values are presented in Table \ref{tab:incidenceomegatable} and can be seen in Figures \ref{fig:dndzdX}, \ref{fig:omegac4values} and \ref{fig:omegasi4values}.

We also compute the column density distribution functions (see Equation \ref{eq:fn}) of \ion{Si}{IV}, \ion{C}{IV} and \ion{Mg}{II} and use a maximum-likelihood estimation (MLE) approach to fit the distributions (see Equation \ref{eq:fnalpha}). The redshift boundaries and column density ranges considered are presented in Table \ref{tab:cddftable} along with the best fit parameters and associated errors. The CDDF distributions and best fits are discussed in Section \ref{sec:cddf} and can be seen in Figures \ref{fig:c4cddf}, \ref{fig:si4c4cddf}, \ref{fig:mg2cddf} and \ref{fig:mg2c4cddf}. Our main findings are:

\begin{enumerate} 

\item {We visually identify 41 \ion{C}{IV} and 7 \ion{Si}{IV} systems with 36 and 7 passing our 5$\sigma$ selection criteria respectively. The highest redshift \ion{C}{IV} and \ion{Si}{IV} absorbers identified in our survey which meet our 5$\sigma$ selection criteria are, respectively, system 15 in sightline $SDSS$ $J1306+0356$ with z = 5.80738 $\pm$ 0.00017 and 9 in sightline $ULAS$ $J0148+0600$ with z = 5.77495 $\pm$ 0.00038. The absorption systems can be seen in {the online appendix}. An example can be see in Figure \ref{fig:U0148b}. }\\

\item{We compute the incidence rates of \ion{C}{IV} and \ion{Si}{IV} and find that both decrease from redshift $\sim$5 to 6. For \ion{C}{IV} we compute $dN/dX$ = 3.6 $\pm$ 0.6 at a median redshift <z> = 4.77 and $dN/dX$ = 0.9 $\pm$ 0.3 at a median redshift <z> = 5.66. For \ion{Si}{IV} we compute $dN/dX$ = 2.2 $\pm$ 1.1 at a median redshift <z> = 5.05 and $dN/dX$ = 0.5 $\pm$ 0.2 at a median redshift <z> = 5.66. The values computed in this work can be seen in Table \ref{tab:incidenceomegatable}. We combine our non comoving incidence rates with those of \citet{DODORICO2013} and \citet{BOKSENBERG2015}. The results can be seen in Figure \ref{fig:dndzdX} and the details are described in Section \ref{sec:incidencerates}.} \\

\item {We compute, for the first time, the comoving mass density of \ion{Si}{IV} ($\Omega_{\ion{Si}{IV}}$) beyond redshift 5.5. We measure $\Omega_{\ion{Si}{IV}}$ = 4.3$^{+2.1}_{-2.1}$ $\times$10$^{-9}$ at <z> = 5.05 and $\Omega_{\ion{Si}{IV}}$ = 1.4$^{+0.6}_{-0.4}$ $\times$10$^{-9}$ at <z> = 5.66. We combine our findings with the values computed in the observational study of \citet{BOKSENBERG2015} and the simulations of \cite{GARCIA2017}. We plot the values in Figure \ref{fig:omegasi4values}. We find that the our $\Omega_{\ion{Si}{IV}}$ values agree very well with the expectations from \cite{GARCIA2017} when considering the column density range of systems identified in this work (12.50 < log(N$_{\textrm{\ion{Si}{IV}}}$) < 14.0). However, when the column density range considered is increased to also include systems with N$_{\textrm{\ion{Si}{IV}}}$<15.0, the expected $\Omega_{\ion{Si}{IV}}$ increases by an order of magnitude. We discuss this in further detail below when comparing the \ion{Si}{IV} CDDFs of this work vs. those of \cite{GARCIA2017}. The associated absorption paths, incidence rates and number of absorbers can be seen in Table \ref{tab:incidenceomegatable}.} \\

\item {We also measure the comoving mass density of \ion{C}{IV} ($\Omega_{\ion{C}{IV}}$) beyond 4.34 and find a similar evolution as previous studies \citep{SIMCOE2006, SIMCOE2011, RYANWEBER2006, BECKER2009, RYANWEBER2009, DODORICO2013}. We combine our results with those from literature and present them in Figure \ref{fig:omegac4values}. We measure $\Omega_{\ion{C}{IV}}$ = 1.6$^{+0.4}_{-0.1}$ $\times$10$^{-8}$ at <z> = 4.77 and $\Omega_{\ion{C}{IV}}$ = 3.4$^{+1.6}_{-1.1}$ $\times$10$^{-9}$ at <z> = 5.66. The associated absorption paths, incidence rates and number of absorbers can be seen in Table \ref{tab:incidenceomegatable}.} \\

\item {We compute the CDDFs of the \ion{Si}{IV} and \ion{C}{IV} systems used in this work as well as the \ion{Mg}{II} systems presented in \cite{CODOREANU2017}. We then perform a MLE functional fit (see Equation \ref{eq:fnalpha}) to the distributions and all best fit parameters can be seen in Table \ref{tab:cddftable}. We find that our \ion{C}{IV} best fit parameters are within 1$\sigma$ of those presented in \cite{DODORICO2013}} \\

\item {We compare our \ion{Si}{IV} and \ion{C}{IV} CDDFs with those presented in and based on the simulations of \cite{GARCIA2017} as can be seen in Figures \ref{fig:si4cddf} and \ref{fig:c4cddfonly}. We find that the \ion{C}{IV} CDDFs are in excellent agreement even beyond the range of \ion{C}{IV} systems identified in our study (log(N$_{\textrm{sys}}$/cm$^{2}$) > 14.25). However, we find that our measured \ion{Si}{IV} CDDF only agrees with that of \cite{GARCIA2017} in the column density range of the systems discovered in this study (12.50 < log(N$_{\textrm{sys}}$/cm$^{2}$) < 14.00). We discuss this in detail in Section \ref{sec:si4discussion}. } \\

\item {We find that 5 of the 7 \ion{Si}{IV} systems have associated \ion{C}{IV} transitions and 10 of the 36 \ion{C}{IV} systems have associated \textit{low$\backslash$high} ionisation systems. We compute the velocity width of the \ion{C}{IV} absorbers ($\Delta$v$_{90}$) and plot them vs. their log(N$_{\textrm{sys}}$/cm$^{2}$) in Figure \ref{fig:c4v90}. We find that all \ion{C}{IV} systems with $\Delta$v$_{90}$ > 200 kms$^{-1}$ have associated \textit{low ionisation} systems. }\\

\item {We investigate the 2 absorbers with the widest velocity widths, systems 8 and 9 in \textit{SDSS J1306 + 0356}. In order to highlight their complex velocity structures and proximity, we plot them on the same velocity scale zeroed on the first component of system 8 in Figure \ref{fig:S1306_z_4d85878}. These systems are also ultra strong \ion{Mg}{II} systems (W$_{2796}$>3 \AA). This class of \ion{Mg}{II} absorbers has been associated with both star formation \citep{NESTOR2011} and tidal interactions \citep{GAUTHIER2013} in group environments below redshift 1. Recently, \cite{CAI2017} image the \textit{SDSS J1306 + 0356} field using HST and find a single LAE with SFR = 2.5 M$_{\sun}$ yr$^{-1}$. Given this, we see it likely that that these systems are tracing a multi-phase medium where hot and cold gas is mixing at the interface between the CGM and IGM. We discuss this in detail in Section \ref{sec:sys8and9discussion}.} \\

\item {We have also identified 5 \ion{Si}{II}, 10 \ion{Al}{II}, 12 \ion{Fe}{II}, 1 \ion{C}{II}, 7 \ion{Mg}{I} and 1 \ion{Ca}{II} transitions associated with \ion{C}{IV}, \ion{Si}{IV} and \ion{Mg}{II} doublets. We compute the associated comoving mass density of each ion and present them in Table \ref{tab:lowions} and Figure \ref{fig:omegavalues}. We compare our values with the work of \cite{WENLAN2017} and find that $\Omega_{\ion{Si}{II}}$, $\Omega_{\ion{Fe}{II}}$, $\Omega_{\ion{Mg}{I}}$ and $\Omega_{\ion{Al}{II}}$ exhibit a flat evolution from redshift 2 to 5. We find that $\Omega_{\ion{Si}{II}}$, $\Omega_{\ion{C}{II}}$ and $\Omega_{\ion{Al}{II}}$ decrease by several orders of magnitude. However, this evolution is due to the association of these systems with only \textit{weak} \ion{Mg}{II} systems beyond redshift 5 while they are mostly associated with $strong$ \ion{Mg}{II} systems below redshift 5. We discuss this in detail in Section \ref{sec:highlow}. } \\

\end{enumerate} 

We find that \ion{C}{IV} systems with log N($_{\textrm{sys}}$) > 14 are preferentially physically associated with \text{low ionisation} systems and are likely to trace multi-phase gas. Further MUSE \citep{BACON2010} observations of these absorbers will be able to identify the physical environment and galaxies responsible for the enrichment and their local contribution to the ionisation field. Recent work by \citet{FUMAGALLI2017} has connected similar absorber pairs to clustered galaxy formation in filamentary structure at z $\simeq$ 3.22. Extending this work to z $\simeq$ 5 provides an exciting opportunity to explore the connection between absorption systems and the associated emission properties of the galaxies responsible for the metal pollution.

\section*{Acknowledgements}
 {We thank Glenn Kacprzak, Michael Murphy, Nikki Nielsen, Stephanie Pointon, Adam Stevens and the anonymous referee for useful discussions and Manodeep Singa for his expertise with code optimisation. This work is based on observations made with the ESO telescopes at the La Silla Paranal Observatory under programme ID 084.A-0390(A). Parts of this research have made use of the Matplotlib library \citep{HUNTER2007}. ERW and AC acknowledge the \textit{Australian Research Council} for \textit{Discovery Project} grant DP1095600 which supported this work. AC is also supported by a Swinburne University Postgraduate Research Award (SUPRA) scholarship. Parts of this research were conducted by the Australian Research Council Centre of Excellence for All-sky Astrophysics (CAASTRO), through project number CE110001020.}


\bibliographystyle{mnras}
\bibliography{paper1_5_biblio} 

\bsp	

\label{lastpage}
\end{document}